\documentclass{article}

\usepackage{PRIMEarxiv}

\usepackage[utf8]{inputenc} 
\usepackage[T1]{fontenc}    
\usepackage{hyperref}       
\usepackage{url}            
\usepackage{booktabs}       
\usepackage{amsfonts}       
\usepackage{nicefrac}       
\usepackage{microtype}      
\usepackage{lipsum}
\usepackage{fancyhdr}       
\usepackage{graphicx}       
\graphicspath{{media/}}     
\usepackage{xcolor}
\usepackage{amsmath}
\DeclareMathOperator*{\argmin}{\arg\!\min}

\title{SDF4CHD: Generative Modeling of Cardiac Anatomies with Congenital Heart Defects \thanks{Code will be open-sourced on GitHub.}
}

\author{
  Fanwei Kong \\
  Department of Pediatrics \\
  Cardiovascular Institute \\
  Stanford University \\
  Stanford\\
  \texttt{fwkong@stanford.edu} \\
   \And
  Sascha Stocker  \\
  Department of Radiology \\
  Stanford University \\
  Stanford \\
  Institute for Biomedical Engineering \\
  ETH Zurich and University Zurich \\
  Zurich \\
  \texttt{sastocke@student.ethz.ch} \\
  \And
  Perry S. Choi \\
  Department of Cardiothoracic Surgery \\
  Stanford University \\
  Stanford \\
  \texttt{pschoi@stanford.edu} \\
   \And
  Michael Ma \\
  Department of Cardiothoracic Surgery \\
  Stanford University \\
  Stanford \\
  \texttt{mma@stanford.edu} \\
  \And
  Daniel B. Ennis \\
  Department of Radiology \\
  Cardiovascular Institute \\
  Stanford University \\
  Stanford \\
  \texttt{dbe@stanford.edu} \\
    \And
  Alison Marsden \\
  Department of Bioengineering\\
  Department of Mechanical Engineering\\
  Department of Pediatrics \\
  Stanford University \\
  Stanford \\
  \texttt{amarsden@stanford.edu} \\
}

\begin{document}
\newcommand{\etal}{\emph{et al.}}
\maketitle

\begin{abstract}
Congenital heart disease (CHD) encompasses a spectrum of cardiovascular structural abnormalities, often requiring customized treatment plans for individual patients. Computational modeling and analysis of these unique cardiac anatomies can improve diagnosis and treatment planning and may ultimately lead to improved outcomes. Deep learning (DL) methods have demonstrated the potential to enable efficient treatment planning by automating cardiac segmentation and mesh construction for patients with normal cardiac anatomies. However, CHDs are often rare, making it challenging to acquire sufficiently large patient cohorts for training such DL models. Generative modeling of cardiac anatomies has the potential to fill this gap via the generation of virtual cohorts; however, prior approaches were largely designed for normal anatomies and cannot readily capture the significant topological variations seen in CHD patients. Therefore, we propose a type- and shape-disentangled generative approach suitable to capture the wide spectrum of cardiac anatomies observed in different CHD types and synthesize differently shaped cardiac anatomies that preserve the unique topology for specific CHD types. Our DL approach represents generic whole heart anatomies with CHD type-specific abnormalities implicitly using signed distance fields (SDF) based on CHD type diagnosis, which conveniently captures divergent anatomical variations across different types and represents meaningful intermediate CHD states. To capture the shape-specific variations, we then learn invertible deformations to morph the learned CHD type-specific anatomies and reconstruct patient-specific shapes. Our approach has the potential to augment the image-segmentation pairs for rarer CHD types for cardiac segmentation and generate cohorts of CHD cardiac meshes for computational simulation. 

\end{abstract}


\section{Introduction}

Congenital heart defects (CHDs) are the most common birth defects affecting 1 in 100 babies and are among the top causes of infant mortality worldwide \cite{ZimmermanSmithSableEchkoWilnerOlsenAtalayAw20}. CHDs are structural abnormalities of the heart or great vessels occurring during fetal development and span a wide array of anatomical characteristics\cite{Liu2019GlobalBP,Marelli2014LifetimePO,Micheletti2018CongenitalHD}. These anatomical abnormalities can dramatically disrupt patients' cardiac function and circulation\cite{Micheletti2018CongenitalHD}. For instance, ventricular septal defects result in abnormal shunting of blood flow between the left and the right ventricles, causing mixing of oxygenated and de-oxygenated blood; d-transposition of the great arteries involves a reversal of the aorta and pulmonary artery connections (i.e., arterioventricular discordance), causing oxygenated blood to be pumped to the body and de-oxygenated blood to be circulated back to the lungs; and pulmonary atresia describes a missing pulmonary outflow tract from the right ventricle, limiting blood flow to the lungs for oxygenation. CHDs form a spectrum of related diagnoses, shaped by variations in the timing, nature, and severity of developmental abnormalities during fetal heart development. CHD defects can appear in isolation or in combination, resulting in unique and complex cardiac malformations across the spectrum.  Approximately 25\% of children born with severe CHD will require heart surgery or other interventions to survive into adulthood \cite{Oster2013TemporalTI}. Given the uniqueness of CHD patients, personalized treatment planning is essential to address their specific needs. 

Personalized modeling of the heart, which encompasses techniques such as 3D printing \cite{Hermsen2020ThreedimensionalPI}, shape analysis \cite{Govil2023BiventricularSM}, and computational simulations \cite{Marsden2015ComputationalMA} holds potential for tailoring surgical and treatment plans to individual patients, ultimately leading to improved outcomes. However, the development and validation of personalized modeling paradigms have mostly been carried out on small patient cohorts, and larger patient cohorts are necessary to validate their clinical advantages. Furthermore, data-driven approaches, such as deep learning methods for segmentation, diagnosis, and shape quantification, are essential to improve the efficiency of personalized modeling, ensuring it can be completed within a reasonable clinical timeframe. Data-driven methods hinge on the availability of large and diverse datasets for training models. However, CHDs, particularly the more complex types (e.g. double outlet right ventricle and hypoplastic left heart syndrome), are rare conditions. This rarity makes it extremely challenging to acquire sufficiently large patient cohorts \cite{Xu2020ImageCHDA3,Fonseca2011TheCA} for validation of new surgical procedures in clinical trials and the development of data-driven approaches for automated image-based computational modeling. Consequently, current developments of surgical treatments for complex CHDs are usually conducted on a very small patient cohort, and there is a lack of consensus among healthcare practitioners regarding the best approaches \cite{Jenkins2002ConsensusbasedMF}. Generative modeling of cardiac anatomies capturing the spectrum of patient variation can generate virtual cohorts that facilitate \emph{in silico} clinical trials \cite{Dou2023ACF,Niederer2020CreationAA} to evaluate new surgical approaches. Namely, generating synthetic CHD models with bespoke abnormalities can significantly expand the available virtual cohort, enabling systematic development and validation of image processing methods, computational modeling approaches, and treatment strategies for any particular type-specific and shape-specific CHD. 

While deep learning (DL) has shown potential in cardiac shape-specific generation \cite{Dou2023ACF,Beetz2022InterpretableCA,Qiao2023CHeartAC,Yuan20234DMR}, prior approaches primarily focused on modeling cardiac structures with normal topology and cannot generalize to CHDs with topologically unusual cardiac structure such as atrial and ventricular septal defects, hypoplastic ventricles, outflow tract atresia, and transposition of the great arteries. Furthermore, prior generative models of cardiac shapes often assumed that the cardiac anatomy distribution had been sufficiently covered in the training dataset. However, CHDs encompass a broad spectrum of rare topological variations among different defect types, making it practically impossible to assemble a comprehensive training cohort from clinical imaging data alone. To address this challenge, it is essential to incorporate prior human knowledge of CHD anatomical abnormalities associated with each type and the relationships between different types into the modeling process. This integration of domain expertise is crucial for enhancing the capacity of these models to accurately represent the diversity of CHD-related cardiac anomalies.

We propose a novel solution, which we term SDF4CHD, to tackle the complex task of representing whole-heart cardiac anatomies for a wide range of CHD types, whether they occur in isolation, in combination, or along a spectrum of disease severity or developmental origin. Namely, our approach enables type-controlled generative modeling of cardiac anatomies, allowing us to create synthetic cardiac anatomies tailored to specific CHD types and conditions. Our key insight is to learn disentangled type and shape representations of CHD hearts. In essence, we represent the \textit{CHD type-specific} heart anatomies and the corresponding abnormalities implicitly using a smooth signed distance function (SDF) learned by Lipschitz-regularized neural networks. This function takes in a CHD diagnosis vector as input and outputs a signed distance field to capture the corresponding unique topologies of complex cardiac anatomies. We learn diffeomorphic deformations to morph the anatomies to capture \textit{shape-specific} anatomical variations in patient cardiac models while preserving the structural anomalies associated with CHDs. A noteworthy advantage of learning smooth SDFs lies in the ability to interpolate between different CHD types, yielding meaningful intermediate representations that are developmentally reasonable, yet may not appear in a sparse dataset. Consequently, following training on a limited CHD dataset, our approach can generate a spectrum of CHD anatomies (types and shapes) that realistically reflects varying degrees of disease severity and anatomy, but which are only represented sparsely in the training dataset. Our key contributions are the following:

\begin{enumerate}
	\item We introduce a cardiac shape representation approach to model the spectrum of CHD. Our innovative method disentangles CHD type and shape using smooth SDFs in conjunction with diffeomorphic deformations. This disentanglement allows us to independently model variations in cardiac topology and shape.
	\item By disentangling type and shape representations, our method enables CHD-type-controlled synthesis of cardiac anatomies and can be used for generating virtual cohorts for rarer CHD types. We demonstrate the applications of our methods by generating both image-segmentation pairs and meshes for computational simulations.
 	\item From a highly heterogeneous CHD dataset, our method can automatically learn spatially-aligned implicit anatomical templates for different types of CHD and represent anatomical abnormalities for both the common and the rare types. It can seamlessly interpolate between binary CHD states and accurately model a spectrum of anatomical changes.
    \item We will openly share our trained CHD whole-heart anatomy generator and the associated virtual cohorts representing various CHD conditions to benefit research in the medical imaging and cardiac computational modeling communities. 
\end{enumerate}

\section{Related Works}
\subsection{Generative Models for Cardiac Anatomies}
Cardiac statistical shape analysis aims to understand how changes in cardiac anatomy relate to diseases, and to enhance disease diagnosis, risk assessment, and treatment strategies. In earlier research on statistical shape models, principal component analysis was commonly used to identify the primary modes of shape variation within a training dataset of cardiac anatomies \cite{Suinesiaputra2018StatisticalSM,Rodero2021LinkingSS,Attar20193DCS}. However, recent advances in deep learning have led to a shift towards employing more expressive variational auto-encoders (VAEs) to model shape variations effectively. For example, Qiao \etal \cite{Qiao2023CHeartAC} proposed a conditional generative model to capture the four-dimensional spatiotemporal anatomy of the biventricular heart conditioned on non-imaging clinical factors. Biffi \etal \cite{Biffi2019ExplainableAS} combined VAE and a classification neural network to model the shape variations of the left ventricle among both healthy individuals and patients with hypertrophic cardiomyopathy. Beetz \etal \cite{Beetz2022InterpretableCA} proposed a mesh VAE to model left ventricle shapes among patients with acute myocardial infarction, and leveraged the latent space of the VAE for predicting major adverse cardiac events. Generative models have also been applied to cardiac image data, enabling the synthesis of new cardiac cine magnetic resonance (MR) images \cite{Duchateau2018ModelBasedGO} and MR images depicting specific pathological conditions \cite{Thermos2021ControllableCS}. Additionally, Campello \etal \cite{Campello2022CardiacAS} proposed an age-conditioned generative adversarial network to generate MR images of aging hearts. 
However, the above studies were all limited to cases in which the heart was topologically normal. Namely, all pathological cases they considered could be modeled by morphing a healthy heart template. As a result, these approaches may not readily extend to CHD patients, whose cardiac anatomy can vary significantly from normal topology (e.g. atrophied chambers, holes, or vascular switches). Furthermore, prior studies mostly focused on the ventricles, yet CHDs encompass anomalies in the atrium and great vessels as well.

\subsection{Patient Image-Based Anatomical Modeling for CHDs}
To the best of our knowledge, no prior study has attempted to automatically model changes in cardiac anatomy across a spectrum of CHDs. Nevertheless, a few studies have concentrated on constructing CHD anatomical models from images for computational simulations \cite{Tikenogullari2023EffectsOC,Vieira2015PatientSpecificIC}, shape analysis \cite{Mauger2021RightleftVS,Govil2023BiventricularSM}, and 3D printing \cite{Xu2019AccurateCH,Loke2017UsageO3,Ryan20183DPF} for specific CHD types. 
For instance, Govil \etal \cite{Govil2023ADL} and Tang \etal \cite{Tang2022ModelingSV} developed mesh templates for specific congenital heart defects, namely tetralogy of Fallot (ToF) and hypoplastic left heart syndrome (HLHS), respectively. These templates were employed to reconstruct patient-specific cardiac meshes from images through template deformation. However, the templates were manually constructed for each specific CHD type and thus may not easily extend to accommodate the multitude of existing CHD types and combinations. Mauger \etal \cite{Mauger2021RightleftVS} and Govil \etal \cite{Govil2023BiventricularSM} utilized principal component analysis (PCA) to analyze biventricular shapes in ToF cases, identifying associations between shape features and factors such as pulmonary regurgitation and the need for pulmonary valve replacements. Similarly, the use of a single ToF mesh template in these studies limits the extension of these methods to other CHD types.  In contrast, our approach is designed to represent a wide range of CHD topologies using a single neural function, thus enabling shape analysis for all CHD types. Loke \etal \cite{Loke2017UsageO3} and Ryan \etal \cite{Ryan20183DPF} reported the beneficial impact of using 3D-printed congenital heart disease models on the learning experiences of clinical residents. However, the 3D-printed anatomies were derived from manually segmenting patient image data, which relies on the availability of patient data and significant user intervention, especially for rare CHD cases. Our method aims to model an augmented range of CHD anatomies by interpolating between various CHD topologies and incorporating shape variations from the training data, resulting in a more diverse set of CHD cardiac anatomies for clinical education.

\subsection{Deep Implicit Template Using SDFs}
An SDF implicitly represents shapes using its zero-level set. DeepSDF introduced an auto-decoder method to approximate a continuous SDF to model shapes with arbitrary topology\cite{Park2019DeepSDFLC}. A few studies have since explored this approach for modeling shapes of cells \cite{Wiesner2022ImplicitNR}, and livers\cite{Raju2021DeepIS}. However, prior methods did not differentiate between type (topology) and shape changes, which can pose challenges when attempting to generate anatomies specific to particular CHDs. In contrast, our study focuses on creating disentangled representations that separate the spectrum of CHD into type-specific and shape-specific dimensions. Our disentangled representation is important for clinical applications, such as generating type-specific generation of synthetic datasets for rare CHD types by varying shapes of the heart while keeping the type fixed, and visualizing treatment scenarios and transition from diseased to healthy states for individual patients by varying the CHD types while keeping the patient-specific shape fixed.

A few studies have proposed to separately model type-specific changes and shape-specific variations by learning implicit templates that model the topology (e.g. type) and another network to deform (e.g. shape) the learned template \cite{Yang2022ImplicitAtlasLD,Sun2022TopologyPreservingSR,Hui2022NeuralTT,Deng2020DeformedIF}. Particularly, Yang \etal\cite{Yang2022ImplicitAtlasLD} proposed to deform an implicit organ template extracted from an SDF to achieve consistent topology in shape reconstruction.
Sun \etal \cite{Sun2022TopologyPreservingSR} proposed to reconstruct meshes of organ anatomies by deforming the meshes extracted from a learned implicit template using Neural Ordinary Differential Equations (NODEs), thereby preventing non-manifold artifacts, including folded faces and self-intersections. However, these approaches either focused on a single template \cite{Yang2022ImplicitAtlasLD, Sun2022TopologyPreservingSR}, or necessitated training distinct networks for different object categories to handle substantial topological changes \cite{Deng2020DeformedIF, Zheng2020DeepIT}. Consequently, the above methods are not readily adaptable to the complexity of CHD anatomies with their diverse and varying topological configurations. 

Hui \etal \cite{Hui2022NeuralTT} suggested a method for learning topology-aware neural templates that are constructed with a set of learned convex shapes to capture topological changes. However, this method only considered general objects (e.g. chairs, tables, airplanes) that can be described by arbitrary composition of basic shape primitives, whereas CHD cardiac anatomies involve both abnormal configurations of multiple cardiac structures and topological variations within a single structure. The modeled topological anomalies should also correspond to meaningful disease states. Our method was thus designed to model the intricate topological variations in multi-structural geometries seen in realistic CHDs. We use CHD diagnosis as an input vector of true or false values corresponding to whether a patient had the specific CHD defects that we chose to include in this study. The inclusion of diagnostic information allows us to reconstruct cardiac geometries that represent the unique cardiac abnormalities associated with each CHD type.

\section{Methods}
\begin{figure}[h]
    \centering
    \includegraphics[width=\linewidth]{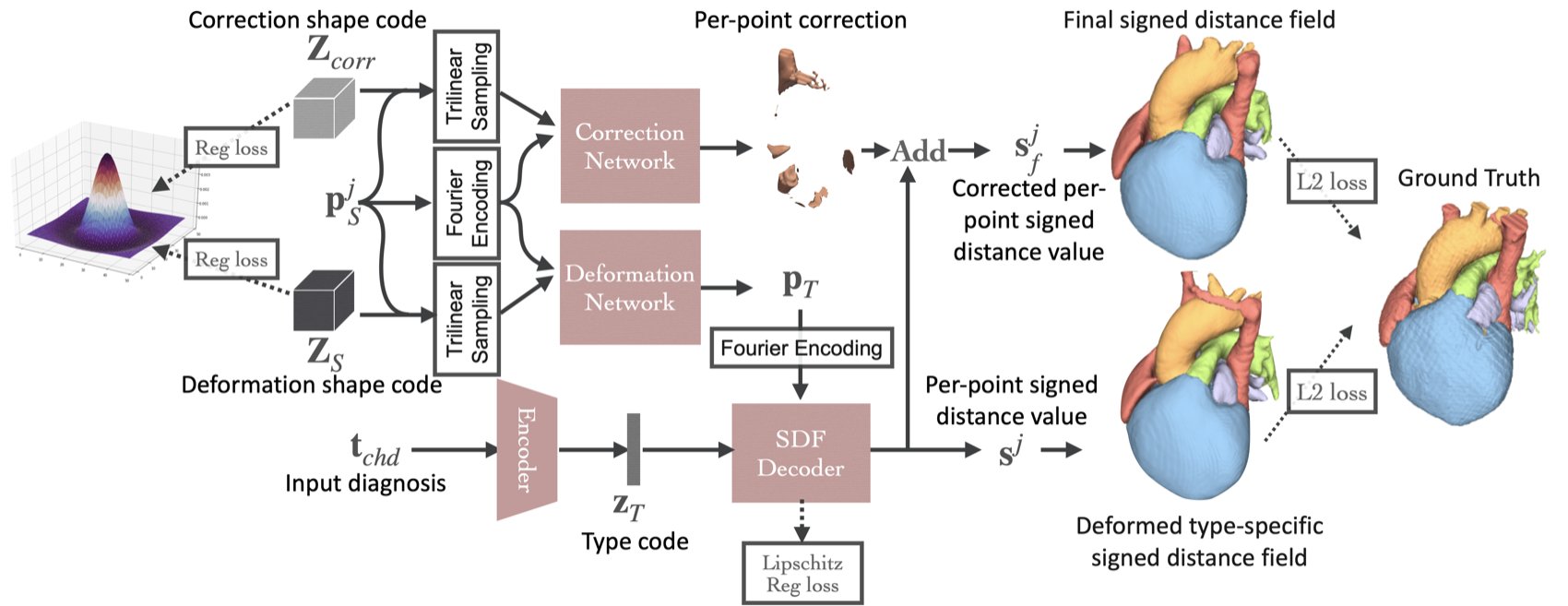}
    \caption{Overview of SDF4CHD. It consists of a type representation module that leverages CHD-type information to predict type-specific cardiac geometries, and a deformation module that changes the shape of the type-specific geometries. Additionally, we used a correction module that captures minor topological variations of patient-specific anatomical features that are not described by their CHD types.}
    \label{fig:method}
\end{figure}

Fig. \ref{fig:method} shows our proposed pipeline, SDF4CHD, for CHD type and shape disentanglement. It consists of a type representation module that leverages CHD-type information to predict type-specific cardiac geometries, a deformation module that changes the shape of the type-specific geometries, and a correction module that captures additional topological variations of patient-specific anatomical features.

\subsection{Type Representation Module}
We modeled the topology changes among and within a total of seven cardiac structures including the myocardium (Myo), blood pools of the right atrium (RA), left atrium (LA), right ventricle (RV) left ventricle (LV), aorta (Ao), and pulmonary arteries (PA).
\subsubsection{Implicit representations for cardiac anatomies} Implicit representations use SDFs to represent 3D geometries of arbitrary topology. An SDF maps a point $\mathbf{p} \in \mathbb{R}^3$ to a scalar value $s \in \mathbb{R}$, indicating the distance from $\mathbf{p}$ to the surfaces of cardiac structures, with the sign denoting whether $\mathbf{p}$ is inside or outside the geometry. However, considering that a heart comprises multiple structures, we opt for modeling a signed distance vector field denoted as $\mathbf{s} = \mathcal{SDF}(\mathbf{p})$. In this context, $\mathbf{s}$ is a vector that represents the signed distances from the domain boundaries of each cardiac structure. Specifically, for a cardiac structure indexed as $i \in {1, \ldots, n}$, if we let $D_i$ represent its domain, $\partial D_i$ represent its boundary, $\bar{D_i}$ represent the closure of $D_i$, $D_i'$ represent the complement of $\bar{D_i}$, and $d$ denote the distance function from $\mathbf{p}$ to $\partial D$, we can express the SDF as follows:

\begin{equation}
s_i = \mathcal{SDF}(\mathbf{p})_i =
    \begin{cases}
    d(\mathbf{p}, \partial D_i), & \text{if } \mathbf{p} \in \bar{D_i} \\
    -d(\mathbf{p}, \partial D_i), & \text{if } \mathbf{p} \in D_i'
\end{cases}.
\end{equation}

The representation of cardiac anatomies is determined by the zero-level set, $\mathcal{SDF}(\cdot) = \mathbf{0}$. 

\subsubsection{Learning type-specific CHD geometries} 

We can parameterize the SDF with an additional input, denoted as $\mathbf{z}_T$, which controls the shape boundaries. In our context, $\mathbf{z}_T$ represents the types of CHDs. This allows the SDF to generate different anatomies corresponding to different types of CHDs. Specifically, we can express it as follows:

\begin{equation}
s_i = \mathcal{SDF}(\mathbf{p}, \mathbf{z}_T)_i=
    \begin{cases}
    d(\mathbf{p}, \partial D_i(\mathbf{z}_T)), & \text{if } \mathbf{p} \in \bar{D_i}(\mathbf{z}_T)\\
    -d(\mathbf{p}, \partial D_i(\mathbf{z}_T)), & \text{if } \mathbf{p} \in D_i(\mathbf{z}_T)'
    \end{cases}, \quad i \in {1, \ldots, 7}.
\end{equation}

Herein, we approximated the above SDFs using Multilayer Perceptrons (MLPs) following \cite{Park2019DeepSDFLC}. Namely,

\begin{equation}
    \mathbf{s} = \mathcal{T}_{dec}(\mathbf{p}, \mathbf{z}_T).
\end{equation}

Since the ReLU activation functions commonly used in MLPs are biased towards low-frequency features, we used Fourier positional encoding to augment the point coordinates $\mathbf{p}$ before using them as inputs \cite{Tancik2020FourierFL}. The final input to $\mathcal{T}_{dec}$ is the product between the point coordinates after the positional encoding and the CHD-type latent code $\mathbf{z}_T$. 

To obtain the CHD-type latent code $\mathbf{z}_T$ from the diagnostic information, we first use another MLP ($\mathcal{T}_{enc}$) to encode the CHD diagnostic information:

\begin{equation}
    \mathbf{z}_T = \mathcal{T}_{enc}(\mathbf{t}_{chd})
\end{equation}

\noindent where the input diagnostic information $\mathbf{t}_{chd}$is a vector containing true or false values, corresponding to whether the patient modeled had the CHD-types found in the training dataset. The CHD-type representation module can thus be expressed as,

\begin{equation}
    \mathbf{s} = \mathcal{T}\left(\mathbf{p}, \mathbf{t}_{chd}\right) = \mathcal{T}_{dec}\left(\mathbf{p}, \mathcal{T}_{enc}\left(\mathbf{t}_{chd}\right)\right).
\end{equation}

The dimension of $\mathbf{t}_{chd}$ is the number of CHD-types modeled. Both $\mathcal{T}_{enc}$ and $\mathcal{T}_{dec}$ consist of six fully connected layers (512 neurons per layer) with leaky ReLU activation functions ($\alpha=0.02$).

\subsection{Diffeomorphic Shape Representation Module}
\subsubsection{Coupling with CHD type representations}
Modeling the shape changes while preserving the cardiac anatomies and topologies specific to CHD types can enable the controlled generation of cardiac shapes for desired CHD types. To preserve the cardiac anatomies learned by the type-specific network, we coupled the type representation module with a conditional shape deformation module $\mathcal{D}$. A shape code $\mathbf{Z}_S^j$ controls how $\mathcal{D}$ deforms the type-specific anatomies to match with each patient-specific anatomy $X^j$ in the training dataset, 

\begin{equation}
    \mathbf{s}^j = \mathcal{T}\left(\mathcal{D}\left(\mathbf{p},\mathbf{z}_S^j\right), \mathbf{t}_{chd}\right).
\end{equation}

Furthermore, we adopted the idea of position-aware shape encoding \cite{Chen2021UNISTUN} where the latent shape code for shape $X^j$ is a latent grid $\mathbf{Z}_S^j$. We linearly interpolated (i.e. $Lerp$) the latent grid at specified points to obtain the point-specific shape code, thus $\mathbf{z}_S^j = Lerp(\mathbf{p}, \mathbf{Z}_S^j)$, where $Lerp(\cdot, \cdot)$ denotes the linear interpolation function. Our spatial latent code $\mathbf{Z}_S^j \in \mathbb{R}^{4\times4\times4\times27}$ had a spatial dimension of four and a latent dimension of 27. We also used Fourier positional encoding \cite{Tancik2020FourierFL} to augment the point coordinates and multiply them with $\mathbf{z}_S^j$. 

\subsubsection{Diffeomorphic flow}
We leveraged the neural ordinary differential equation (NODE) approach \cite{Chen2018NeuralOD,Hui2022NeuralTT,Sun2022TopologyPreservingSR} to learn an invertible deformation between the learned type-specific geometries (type space) and patient-specific cardiac geometries (shape space). Namely, the diffeomorphic flow $\phi(\mathbf{p}, \mathbf{z}_S^j, t)$ of patient anatomy $X^j$ is the solution to the initial value problem of an ODE and describes the trajectory of points moving from their locations $\mathbf{p}_S \in \mathbb{R}^3$ on the signed distance field of patient anatomy $X^j$ at $t=0$ to their corresponding locations $\mathbf{p}_T \in \mathbb{R}^3$ on the signed distance field of the CHD type-specific template at $t=T$, so that

\begin{equation}
    \frac{\partial \phi(\mathbf{p}, \mathbf{z}_S^j, t)}{\partial t} = \mathbf{v}(\phi(\mathbf{p}, \mathbf{z}_S^j,  t), \mathbf{z}_S^j, t) \quad \text{s.t.} \quad \phi(\mathbf{p}, \mathbf{z}_S^j, 0) = \mathbf{p}_S^j \quad \text{and} \quad \phi(\mathbf{p}, \mathbf{z}_S^j, T) = \mathbf{p}_T.
    \label{eq:ode}
\end{equation}

The unique solution of Equation \ref{eq:ode} exists if $\mathbf{v}(\cdot, \cdot, \cdot)$ is Lipschitz continuous. Namely, there exists $L\in\mathbb{R}^+$ such that, for any $\mathbf{x}, \mathbf{y} \in \mathbb{R}^3$,

\begin{equation}
    ||\mathbf{v}(\mathbf{x}) - \mathbf{v}(\mathbf{y}) ||_p \leq L || \mathbf{x} - \mathbf{y} ||_p,
\end{equation}

\noindent where $||\cdot||_p $ is the $p-$norm for $p\geq 1$. Our deformation module approximates $\mathbf{v}(\cdot, \cdot, \cdot)$ using six fully connected layers with leaky ReLU activation functions ($\alpha=0.02$), and is thus differentiable and Lipschitz continuous \cite{Ma2022CortexODELC}. Therefore, the trajectories of different points will, in theory, not intersect with each other. This ensures that flowing mesh points from the type space to the shape space will not generate intersecting mesh surfaces. This ODE can be integrated to progressively map point locations from the shape space to the type space. The diffeomorphic deformation conditioned on the deformation code is thus given by,

\begin{equation}
    \mathbf{p}_T = \mathcal{D}(\mathbf{p}, \mathbf{z}_S^j) = \phi(\mathbf{p}, \mathbf{z}_S^j, T) = \mathbf{p}_S^j + \int_0^T \mathbf{v}(\mathbf{z}_S^j, \phi(\mathbf{p}, \mathbf{z}_S^j, t), t) dt.
    \label{eq:forward}
\end{equation}

By the existence and uniqueness properties of ODEs, the NODE nets are invertible. Therefore, we can readily obtain the inverse deformation by integrating the ODE backward in time:

\begin{equation}
  \mathbf{p}_S^j = -\mathcal{D}(\mathbf{p}, \mathbf{z}_S^j) = \psi(\mathbf{p}, \mathbf{z}_S^j, T) = \mathbf{p}_T + \int_0^T -\mathbf{v}(\mathbf{z}_S^j, \phi(\mathbf{p}, \mathbf{z}_S^j, t), t) dt,
  \label{eq:backward}
\end{equation}

where $\psi:\mathbb{R}^3 \times [0, T] \rightarrow \mathbb{R}^3$ is the inverse of the diffeomorphic flow $\phi$. We used the forward Euler method to solve the ODE in Equations \ref{eq:forward} and \ref{eq:backward}, with a fixed step size of 0.2 over a time interval of $t \in [0, 1]$.

\subsection{Correction Module} 
CHD cardiac geometries are primarily described by deformations from CHD type-specific templates. However, additional topological variations may exist among patients with the same CHD type (e.g. different great vessel connections). We thus use an additional MLP ($\mathcal{C}$) to output a correction vector $\mathbf{c} \in \mathbb{R}^7$ at each point $\mathbf{p}_S$ in the shape space:

\begin{equation}
    \mathbf{c}^j = \mathcal{C}(\mathbf{p}, \mathbf{z}_C^j).
\end{equation}

Similar to the deformation module, we use positional-aware shape encoding by sampling the correction code from a latent grid $\mathbf{Z}^j_c \in \mathbb{R}^{4\times4\times4\times27}$, using trilinear interpolation $\mathbf{z}_C^j = Lerp(\mathbf{p}, \mathbf{Z}_c^j)$. This correction describes the minor variations of cardiac anatomies at the point locations that cannot be modeled by simply deforming type-specific cardiac shapes. Therefore, the final SDF prediction $\mathbf{s}_f^j$, for a point $\mathbf{p}_S^j$ in the shape space of patient anatomy $X^j$ is, 

\begin{equation}
    \mathbf{s}_f^j = \mathcal{T}\left(\mathcal{D}\left(\mathbf{p}_S^j, \mathbf{z}_S^j\right), \mathbf{t}_{chd}\right) + \mathcal{C}(\mathbf{p}_S^j, \mathbf{z}_C^j), \quad \text{where, } \mathbf{z}_C^j = Lerp(\mathbf{p}, \mathbf{Z}_c^j), \quad \mathbf{z}_S^j = Lerp(\mathbf{p}, \mathbf{Z}_S^j).
\end{equation}

The correction network contains six fully connected layers with 512 neurons per layer, connected with leaky ReLU activation function ($\alpha=0.02$). Similar to the type and the shape representation modules, we also used Fourier positional encoding \cite{Tancik2020FourierFL} to augment the point coordinates and multiplied them with $\mathbf{z}_C$. 

\subsection{Optimization}
Leveraging the ground truth anatomies obtained from expert annotations and the corresponding diagnosis information, we can jointly learn the generic CHD type representations and the posterior probability distribution of the shape and correction codes that best represent the patient-specific anatomies. Assuming a prior distribution of the shape codes, we can then sample a shape code from this prior distribution to construct synthetic cardiac anatomies for a specific CHD type beyond those included in the training dataset. We followed the DeepSDF auto-decoder approach described by Park \etal \cite{Park2019DeepSDFLC}, where we jointly optimized the network parameter, the shape, and the correction codes to maximize the joint log posterior over all training shapes. Namely, given a training dataset of CHD patient anatomies $\{X^j\}_{j=1}^N$, we optimize the network and latent codes over a set of point coordinates and their ground truth signed distance values so that:

\begin{equation}
    X^j = \{(\mathbf{p}_S^j, \mathbf{s}_{gt}^j) : \mathbf{s}_{gt}^j = \mathcal{SDF}^j(\mathbf{p}_S^j)\}.
\end{equation}

If we assume that the deformation needed to model patient-specific geometries and the need for additional corrections are independent events, the posterior probability of shape and correction codes given the training data pairs can be expressed as:

\begin{equation}
    p_\theta(\mathbf{Z}^j_S, \mathbf{Z}^j_C | X^j) = p(\mathbf{Z}^j_S)p(\mathbf{Z}^j_C) \prod_{(\mathbf{p}_S^j, \mathbf{s}_f^j)\in X^j} p_\theta(\mathbf{s}_f^j|\mathbf{Z}^j_S, \mathbf{Z}^j_C , X^j).
    \label{eq:posterior}
\end{equation}

We assumed the prior distribution over the shape codes and the correction codes to be zero-mean multivariate-Gaussian distributions, $\mathcal{N}(0, \sigma^2I)$. The likelihood $p_\theta(\mathbf{s}_f^j|\mathbf{Z}^j_S, \mathbf{Z}^j_C , X^j)$ is assumed to be 

\begin{equation}
p_\theta(\mathbf{s}_f^j|\mathbf{Z}^j_S, \mathbf{Z}^j_C , X^j) = \exp\left(-\mathcal{L}_{geo}(\mathbf{s}_f^j, \mathbf{s}_{gt}^j) - \mathcal{L}_{geo}(\mathbf{s}^j, \mathbf{s}_{gt}^j)\right),
\end{equation}
 where $\mathcal{L}_{geo}$ is the loss function responsible for the accuracy of the predicted signed distance values both after the shape deformation module and after the correction module. In practice, we found that training with occupancy values (i.e. $0$ if the signed distance value is negative and $1$ if otherwise) resulted in faster convergence than training with signed distance values. We used the L2 norm as the reconstruction loss function, namely, $\mathcal{L}_{geo}(\mathbf{s}_f^j, \mathbf{s}_{gt}^j) = ||\mathbf{s}_f^j-\mathbf{s}_{gt}^j||_2^2$, and $\mathcal{L}_{geo}(\mathbf{s}^j, \mathbf{s}_{gt}^j) = ||\mathbf{s}^j -\mathbf{s}_{gt}^j||_2^2$. Therefore, during training, we first sample points near the surfaces of the heart. At these points, we optimize both the network parameters $\theta$, the individual shape codes $\{\mathbf{Z}^j_S\}_{j=1}^N$ , and correction codes $\{\mathbf{Z}^j_C\}_{j=1}^N$ to maximize the joint $log$ posterior in Equation \ref{eq:posterior}. The objective function is as follows,
 
\begin{equation}
    \argmin_{\theta, \{\mathbf{Z}^j_S\}_{j=1}^N, \{\mathbf{Z}^j_C\}_{j=1}^N} \sum_{j=1}^N \left( \mathcal{L}_{geo}(\mathbf{s}_f^j, \mathbf{s}_{gt}^j) + \mathcal{L}_{geo}(\mathbf{s}^j, \mathbf{s}_{gt}^j) + \frac{1}{\sigma^2} ||\mathbf{Z}_S||_F + \frac{1}{\sigma^2} ||\mathbf{Z}_c||_F\right).
\end{equation}

To further encourage smooth latent spaces learned by the type representation module, we applied Lipschitz regularization \cite{Liu2022LearningSN} on $\mathcal{T}_{dec}$ by penalizing the upper bound of its Lipschitz constant. The regularization loss is $\mathcal{L}_{lip} = \prod^L_k softplus(l_k)$, where $l_k = ||W_k||_p$ is the norm of the weight matrix $W_k$ at layer $k$, and $softplus(l_k) = ln(1+e^{l_k})$ to ensure positive Lipschitz bounds. We normalized the weight matrix $W_k$ with $l_k$, and added $\mathcal{L}_{lip}$ to the objective function to bound the Lipschitz constant of $\mathcal{T}_{dec}$, namely, 

\begin{equation}
    \argmin_{\theta, \{\mathbf{Z}^j_S\}_{j=1}^N, \{\mathbf{Z}^j_C\}_{j=1}^N} \sum_{j=1}^N \left( \mathcal{L}_{geo}(\mathbf{s}_f^j, \mathbf{s}_{gt}^j) + \mathcal{L}_{geo}(\mathbf{s}^j, \mathbf{s}_{gt}^j) + \frac{1}{\sigma^2} ||\mathbf{Z}_S||_F + \frac{1}{\sigma^2} ||\mathbf{Z}_c||_F + \lambda_{lip} \mathcal{L}_{lip}\right).
\end{equation}

We note that the shape representation module requires learned CHD type-specific implicit templates to reconstruct shapes in the dataset. However, during the early stages of training, the type representation module has yet to learn accurate representations of CHD type-specific anatomies. Consequently, it is unable to provide reliable initial templates for the shape representation module to deform. Therefore, we initialize the type representation module by pre-training it to reconstruct healthy cardiac anatomy regardless of the input diagnosis vector $\mathbf{t}_{chd}$ for 500 iterations of gradient descent updates. Furthermore, at the start of training, jointly optimizing the type and the shape representation modules may interfere with the training of both. Therefore, we alternatively train and freeze the type representation module and other parts of the networks every five epochs for the first 1000 epochs and jointly optimize all network parameters for the remaining 1000 epochs. 

During inference, we fix the optimized network parameters $\theta$ and estimate the shape and the correction codes via Maximum-a-Posterior:

\begin{equation}
    \hat{\mathbf{Z}_S}, \hat{\mathbf{Z}_C} = \argmin_{\mathbf{Z}_S, \mathbf{Z}_C} \sum_{(\mathbf{p}^j, \mathbf{s}_{gt}^j)} \left( \mathcal{L}_{geo}(\mathbf{s}_f^j, \mathbf{s}_{gt}^j) + \mathcal{L}_{geo}(\mathbf{s}^j, \mathbf{s}_{gt}^j) + \frac{1}{\sigma^2} ||\mathbf{Z}_S||_F + \frac{1}{\sigma^2} ||\mathbf{Z}_C||_F\right).
    \label{eq:total}
\end{equation}

We initialize the shape and correction codes randomly from $\mathcal{N}(0, 0.01^2I)$, and optimize the above objective function for 200 iterations using the Adam optimizer to reconstruct unseen anatomies.

\section{Experiments and Results}
\subsection{Datasets}
In our study, we considered several CHD types that cause abnormal anatomical configurations of the heart chambers and great vessels: ventricular septal defects (VSD), atrial septal defects (ASD), Tetralogy of Fallot (ToF), d-transposition of the great arteries (TGA), double outlet right ventricle (DORV), and pulmonary atresia (PuA). VSDs were defined as isolated defects of the ventricular septum, without further subcategorization. ToF was defined as the classic constellation of four associated defects: anterior maligned VSD, right ventricular outflow tract obstruction, right ventricular hypertrophy, and overriding aorta. TGA was defined as anterior-ventricular discordance in the absence of atrioventricular discordance and lack of a subpulmonary conus. DORV was defined as any defect with the presence of a VSD and both subpulmonary and subaortic conus. PuA was defined as discontinuity of the pulmonary artery, regardless of other lesions present. The training and testing data comes from the public “imageCHD” dataset\cite{Xu2020ImageCHDA3} that includes 110 CT images, the corresponding diagnosis (CHD type) and ground truth segmentations of seven cardiac structures. Since the ``imageCHD” dataset was initially designed for classification, the provided ground truth segmentations were extremely noisy, posing a significant challenge in accurately differentiating the true defects from noise. Therefore, we manually processed the segmentations under the supervision of a radiology expert to remove noise and improve consistency with diagnostic information (Please see Supplementary Materials for a visual comparison). 

\begin{figure}[h]
    \centering
    \includegraphics[width=\linewidth]{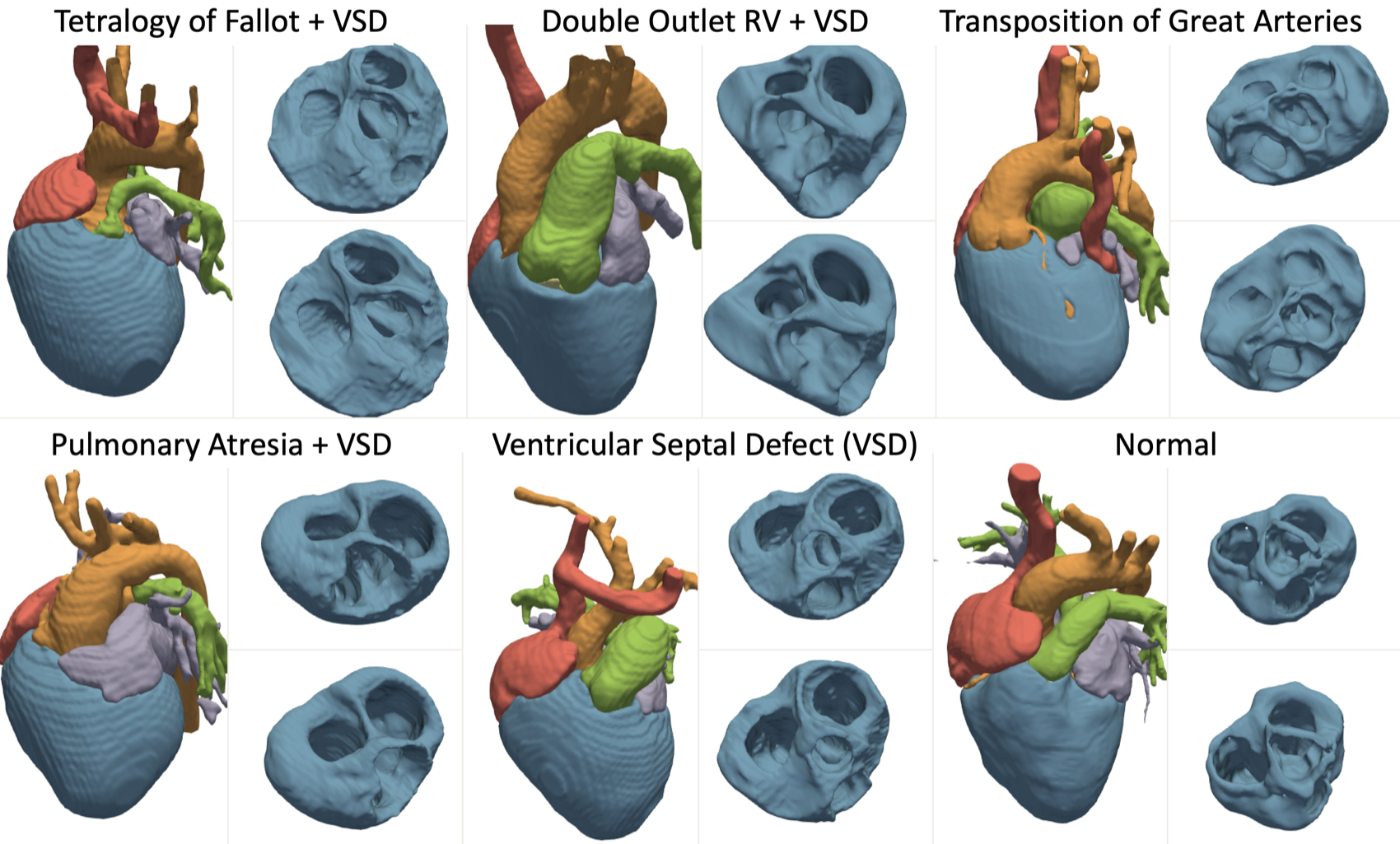}
    \caption{Example ground truth whole heart anatomies for various CHD types.}
    \label{fig:dataset}
\end{figure}

\subsection{Baselines}
The key insights of our SDF4CHD approach include: 1) disentangling the CHD-type and CHD-shape representation, and 2) leveraging the CHD diagnostic information and neural SDFs to represent topological variations across CHD types. We then verify the advantages of our approach by comparing it with the following baselines. First, we adapted DeepSDF \cite{Park2019DeepSDFLC}, an auto-decoder approach to represent single-category objects, to represent multiple cardiac structures by predicting a signed distance vector field. Then, to verify the advantage of using a disentangled representation between CHD-type and CHD-shape, we implemented a conditional DeepSDF (CDeepSDF) that directly predicts the SDF point values based on the point coordinates concatenated with the CHD-type and CHD-shape latent codes. We also compared with Neural Diffeomorphic Flow (NDF) \cite{Sun2022TopologyPreservingSR}, a prior approach that applied implicit templates to medical applications but only considered a single implicit template with consistent topology. We compare our SDF4CHD method with this approach to verify the necessity of allowing topological variations in modeling CHD anatomies. In all these baselines, the latent codes are obtained in the same manner as our method, ensuring a fair comparison. Namely, the same dimension, regularization, and trilinear sampling method were applied for all baselines. We also used the same point sampling method during both training and testing to construct the signed distance field predictions for all methods. Please see Supplementary Materials for further implementation details of baselines. 

\subsection{Learning the generic cardiac anatomies for various CHDs}
\begin{figure}[h]
    \centering
    \includegraphics[width=\linewidth]{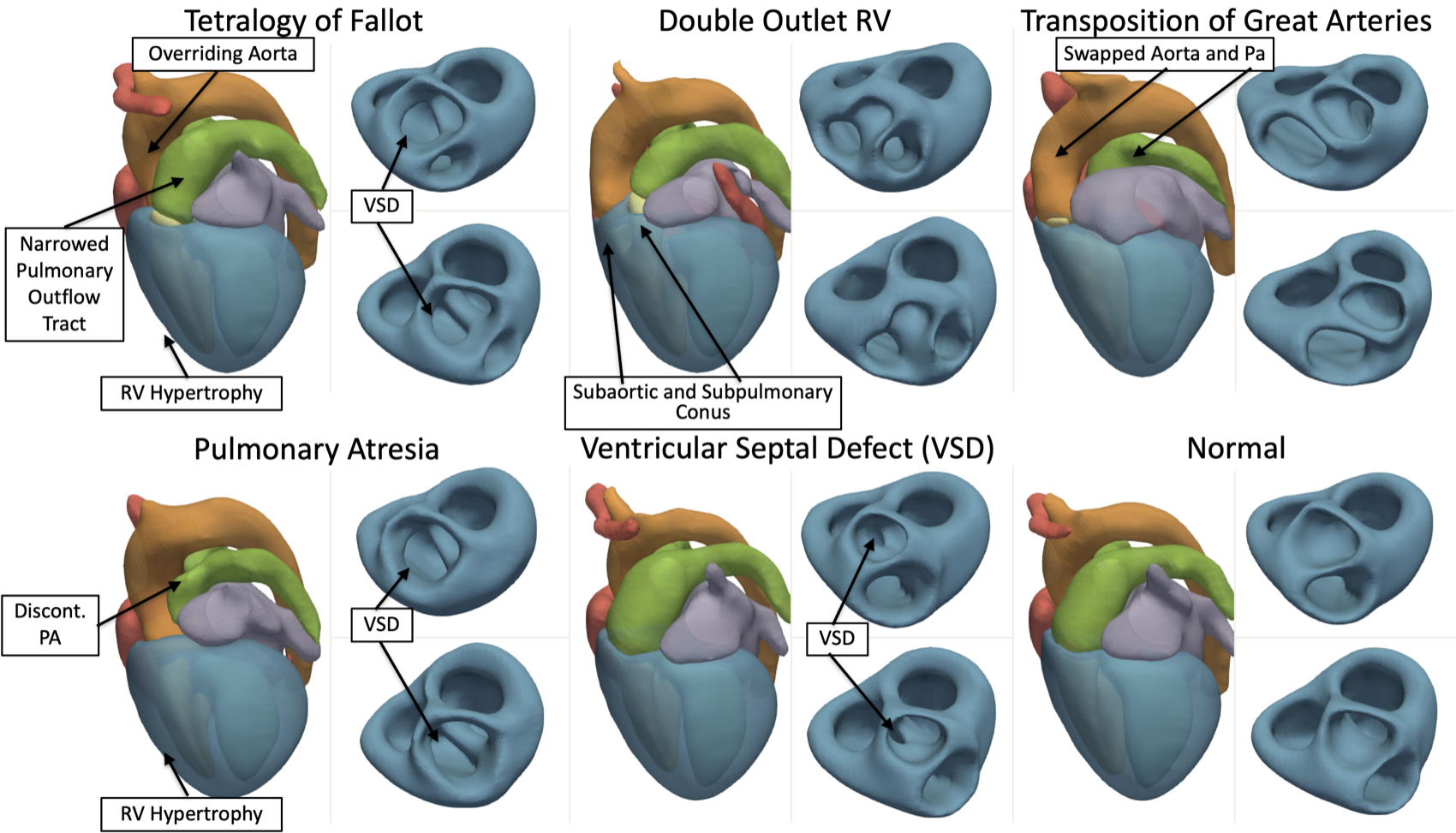}
    \caption{Cardiac geometries extracted from the learned type-specific SDFs for different CHDs.}
    \label{fig:type}
\end{figure}

We first examined the learned type-specific anatomical templates. As shown in Figure \ref{fig:type}, our method successfully learned type-specific geometries that represent the typical abnormalities associated with each CHD type. For instance, it learned to represent a hole in the interventricular septum for VSD, the discontinuation of the pulmonary artery in PuA, the transposition of the aorta and pulmonary positions in TGA, the presence of VSD and both subpulmonary and subaortic conus in DORV, and the existence of VSD, overriding aorta, right ventricular hypertrophy, and stenosis in the pulmonary outflow tract for ToF. We note that our method captured both the topological differences in CHD anatomies and the characteristic shape abnormalities specific to each CHD type when compared to healthy anatomy. For example, features such as right ventricle hypertrophy and a narrowed pulmonary outflow tract represent shape abnormalities rather than changes in topology. Therefore, the learned implicit type templates encompass both the topology and shape abnormalities unique to each CHD type.

\subsection{Modeling the intermediate CHD states.}

During our training process, we used a vector to represent CHD diagnosis information, indicating whether a patient had specific CHD types through binary values. However, CHDs form a spectrum of related diagnoses shaped by
variations in the timing, nature, and severity of developmental abnormalities during fetal heart development. We show that our learned type latent space has demonstrated the ability to encompass intermediate states between different CHD types, effectively representing a spectrum of CHDs.

\begin{figure}[h]
    \centering
    \includegraphics[width=\linewidth]{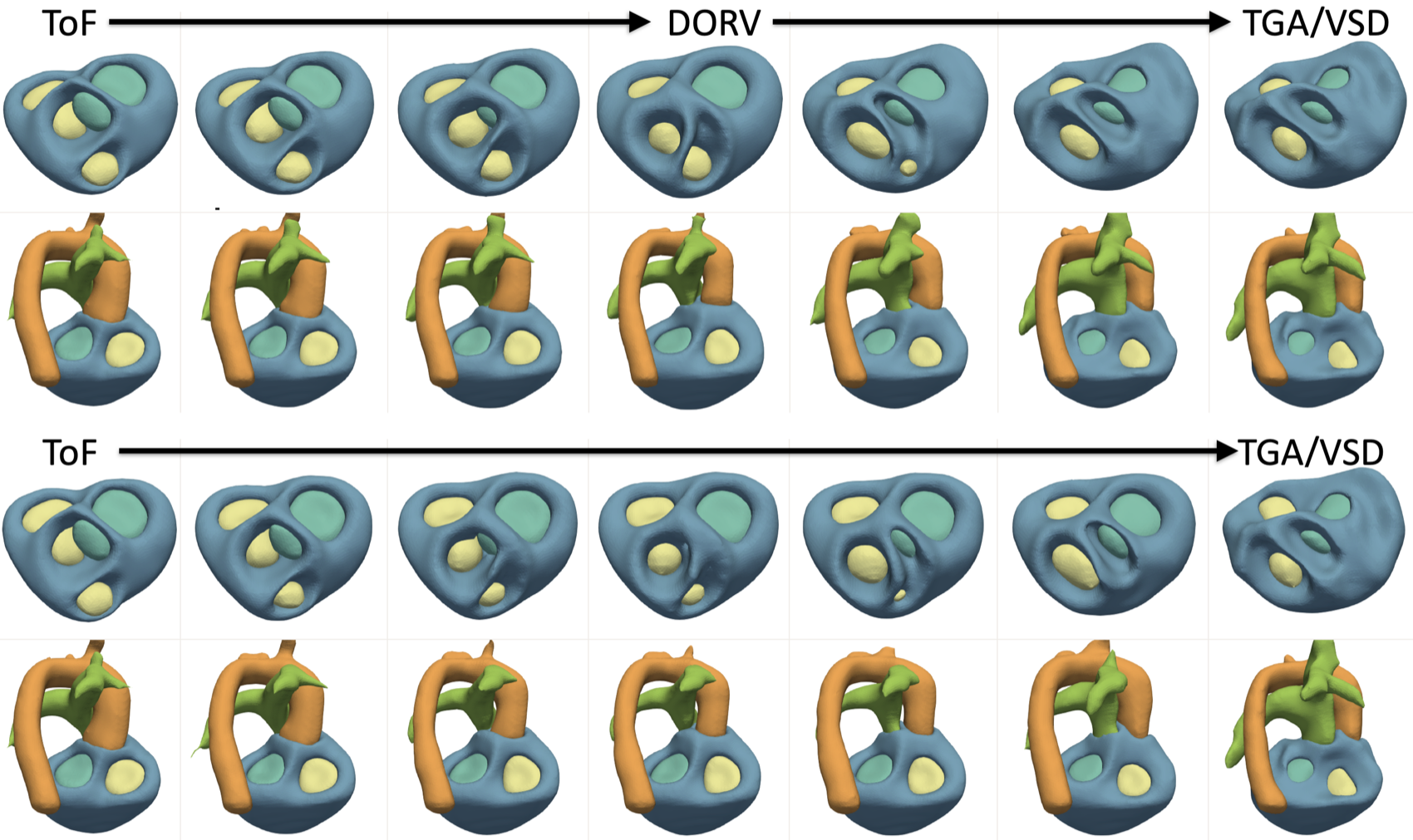}
    \caption{Interpolated cardiac anatomies between ToF and TGA (with VSD) to illustrate that our learned latent space can cover a spectrum of CHD diseases. Top: interpolating the diagnosis vectors from ToF to DORV and from DORV to TGA. Bottom: interpolating the diagnosis vectors directly from ToF to TGA}
    \label{fig:interp_diag_dorv}
\end{figure}

For instance, CHD conditions of DORV, ToF, and d-TGA all fall under the category of conotruncal defects. These defects can be thought of as existing along a spectrum of conal septum distribution. In simpler terms, DORV, characterized by the presence of both a subaortic and subpulmonic conus, can be seen as occupying a position between two endpoints: ToF (where there is a subpulmonary conus and no subaortic conus) and d-TGA (where there is a subaortic conus and no subpulmonic conus). The variation in conal distribution is responsible for the differences in the rotational positions of the great arteries across these three defects. Indeed, it is this spectrum of conal septum distribution that makes these diagnoses (ToF, DORV, and d-TGA) an ideal template for testing our method's capability to represent intermediate states along a spectrum. Figure \ref{fig:interp_diag_dorv} (top) illustrates the intermediate CHD anatomies from a ToF diagnosis to a DORV diagnosis, and from the DORV diagnosis to a dTGA diagnosis, showing rotating positions of the aorta and pulmonary arteries. We note that our method did not require the intermediate diagnosis DORV vector to represent this spectrum. Namely, interpolating directly between a ToF diagnosis vector and a TGA diagnosis vector can represent a similar transition in cardiac anatomies. This shows that our method can readily represent meaningful intermediate states that correspond to intermediate diagnosis. 

\begin{figure}[h]
    \centering
    \includegraphics[width=\linewidth]{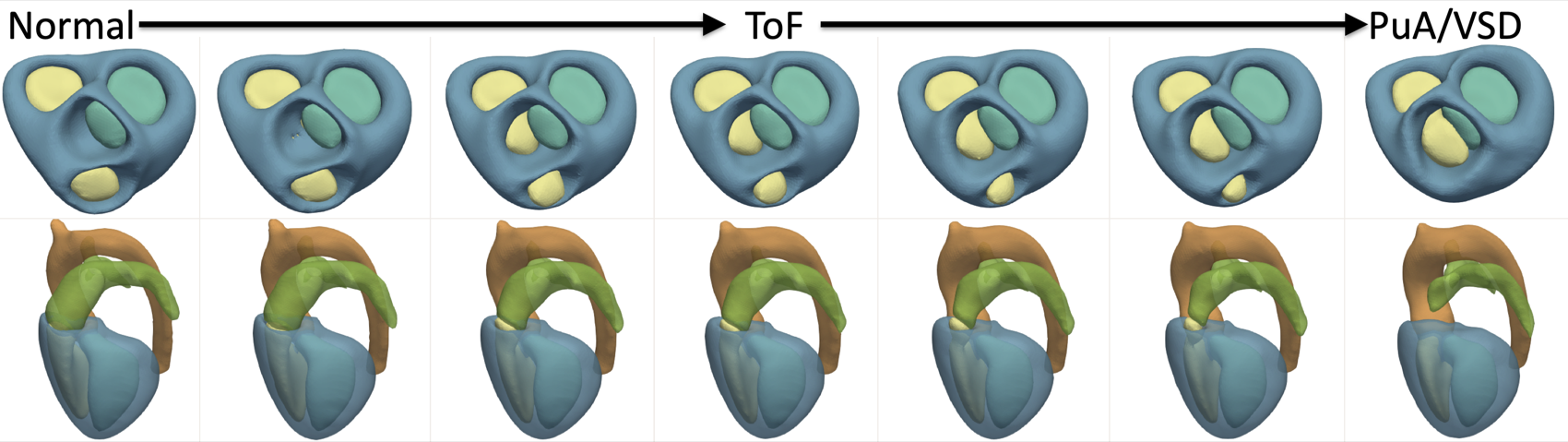}
    \caption{Interpolated cardiac anatomies between normal, ToF and PuA (with VSD) to illustrate that our learned latent space can continuously describe different severities of CHD disease.}
    \label{fig:interp_diag_tof}
\end{figure}

Similarly, our method can also encompass intermediate states along a spectrum of disease severity. For instance, the symptoms of ToF can vary significantly from one individual to another. The severity of these symptoms, which can range from mild to severe, is linked to the extent of blood flow obstruction originating from the right ventricle to the lungs. Pulmonary atresia with VSD is often classified as a severe form of ToF, characterized by the complete constriction of the pulmonary outflow tract. Figure \ref{fig:interp_diag_tof} (top) illustrates the intermediate CHD anatomies from normal to a ToF diagnosis and from the ToF diagnosis to a PuA (with VSD) diagnosis. Along this spectrum, our method captured the forming of a large VSD, increasingly overriding aorta, thickening of RV myocardium, and the gradual narrowing and ultimate detachment of the pulmonary artery from the right ventricle.

\subsection{Disentangled representation of CHD types and shapes}

\begin{figure}[h!]
    \centering
    \includegraphics[width=\linewidth]{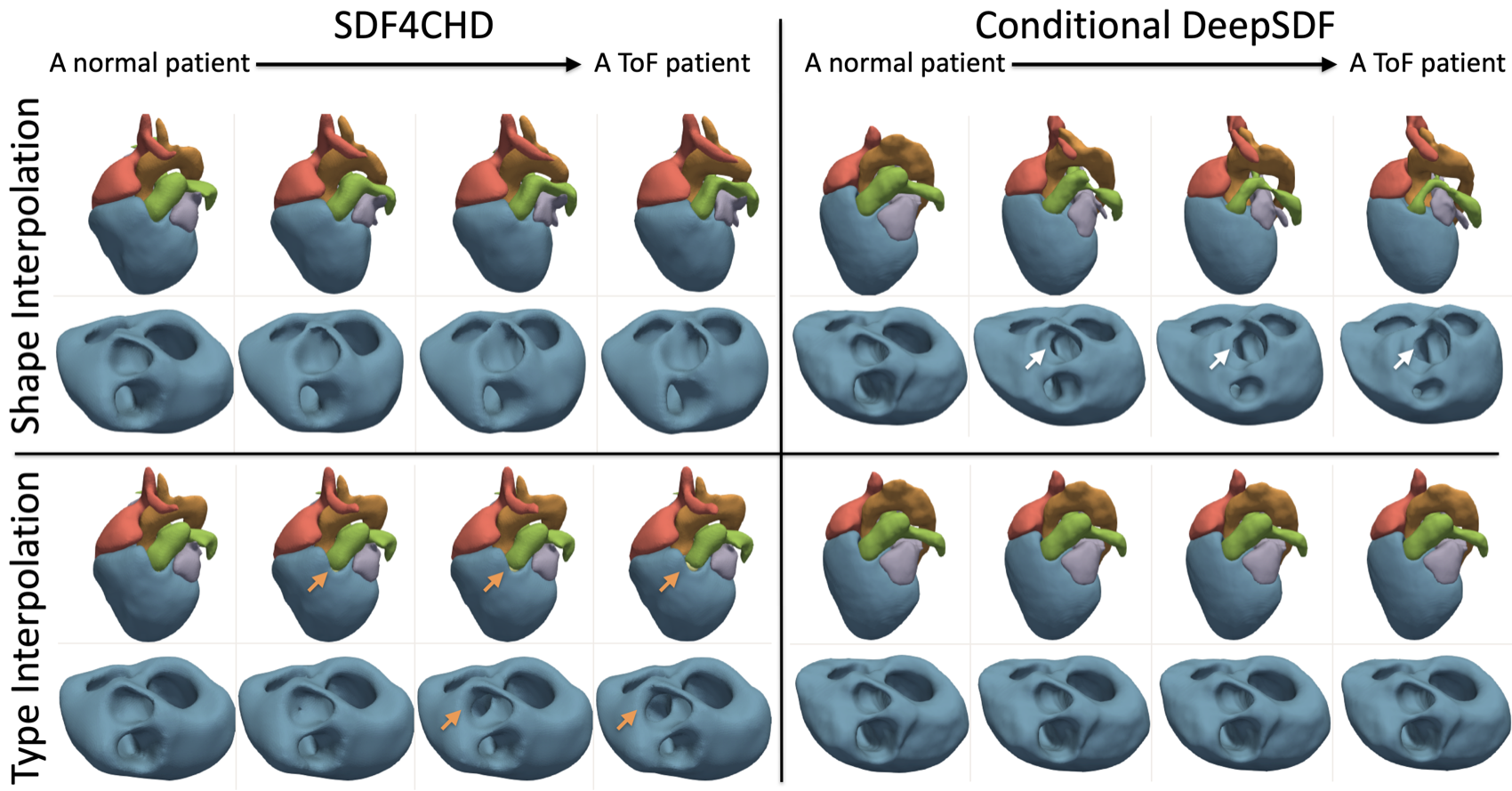}
    \caption{Reconstruction results when interpolating the latent type or shape code. The top panels compare the anatomical changes achieved by our approach and conditional DeepSDF when keeping the CHD type code fixed and interpolating the shape codes from a normal patient to a ToF patient. In the bottom panels, we compare the anatomical changes when keeping the shape codes fixed and interpolating the CHD type codes from normal to ToF. The orange arrows highlight that our method produced the expected anatomical changes from normal to ToF, whereas the white arrows highlight that conditional DeepSDF resulted in undesired type changes when the type code was fixed and failed to capture type changes when changing the type code.}
    \label{fig:interp_compare}
\end{figure}

A major advantage of disentangling the learned shape and type representations is that we can then manipulate the learned shape and type independently. Figure \ref{fig:interp_compare} the results of linearly interpolating the shape latent space or type latent space between a patient with normal heart chambers and a ToF patient. When fixing the shape latent code and interpolating type code from normal to ToF, the VSD and narrowed PA featured in ToF gradually appear, whereas the overall shape of the heart remains roughly fixed. When fixing the type latent code and interpolating the shape code, our reconstruction preserved the type-specific anatomies while only deforming the shape. In contrast, the baseline method, conditional DeepSDF, lacked the ability to separately manipulate learned type and shape anatomies. Specifically, during shape interpolation with a fixed type latent code, the topology of the resulting anatomies underwent alterations. Likewise, when interpolating the type code while maintaining a fixed shape code, the resulting anatomical structures did not accurately reflect the expected topology changes from ToF to normal anatomies.

\begin{figure}[h!]
    \centering
    \includegraphics[width=\linewidth]{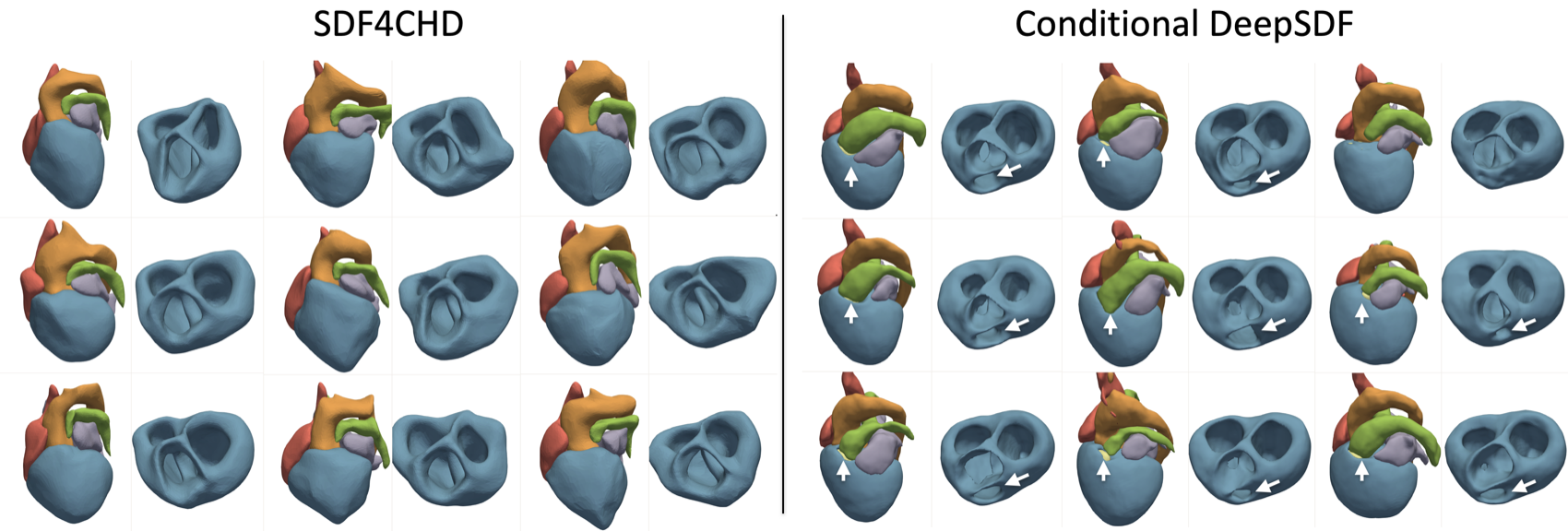}
    \caption{Our approach (left) readily enables type-controlled generation for pulmonary atresia with VSD (PuA+VSD). In contrast, conditional DeepSDF struggled to learn the correct cardiac topology when conditioned on the CHD diagnosis, resulting in numerous topological errors, as highlighted by the white arrows.}
    \label{fig:gen}
\end{figure}

We can thus apply our method to generate synthetic anatomies for rare CHD types, augmenting the amount of available training data for these specific CHDs. For example, PuA with VSD is a rare condition with an incidence of 7 per 100,000 live births \cite{Naimi2021AccuracyOF}. Both acquiring sufficient data for this CHD type and manually segmenting these data would be challenging. To generate new anatomies with this specific CHD type, we fixed the input type diagnosis vector and varied the input shape code to the learned PuA (with VSD) type template. Namely, the shape code was obtained by randomly sampling from the Gaussian prior distribution. We applied the same treatments to the conditional DeepSDF for comparison. The left panel of figure \ref{fig:gen} shows that the example cardiac anatomies generated by our method successfully preserve the correct cardiac anatomy of PuA (with VSD), whereas the right panel shows that the anatomies generated by the conditional DeepSDF failed to preserve the topology.
 
\subsection{Quality of CHD Shape Representation and Reconstruction}

\begin{table}[h!]
\caption{Dice scores for whole heart (WH) and individual cardiac structures in the represented seen geometries compared to different baselines.(* denotes significance differences in mean values when compared with ``SDF4CHD''; denotes significance differences in mean values when compared with ``SDF4CHD+Corr''; we assume p-Values$< 0.0125 = 0.05/4$, based on a Bonferroni Correction for 4 comparisons.)}
\label{table:seen}
\resizebox{\textwidth}{!}{%
\begin{tabular}{lllllllll}
\toprule
{} &              WH &              LV &              RV &              LA &             RA &             Myo &              Ao &               PA \\
\midrule
SDF4CHD      &    89.7$\pm$1.5 &    94.3$\pm$2.2 &    92.5$\pm$2.2 &    86.3$\pm$3.9 &   90.4$\pm$2.8 &    90.7$\pm$3.1 &    84.4$\pm$7.4 &    75.5$\pm$14.3 \\
SDF4CHD+Corr &   91.8$\pm$1.0* &    94.3$\pm$2.1 &   92.6$\pm$2.2* &   90.2$\pm$2.4* &  93.5$\pm$1.7* &   91.5$\pm$2.6* &   89.2$\pm$4.1* &    82.7$\pm$8.5* \\
NDF       &  90.1$\pm$1.5*! &    94.2$\pm$2.4 &    92.4$\pm$2.0 &  87.9$\pm$3.9*! &  90.8$\pm$2.8! &    91.0$\pm$2.7 &   84.6$\pm$7.2! &  77.1$\pm$13.7*! \\
CDeepSDF  &  92.1$\pm$0.7*! &  95.5$\pm$1.2*! &  94.5$\pm$1.3*! &   89.9$\pm$2.5* &  93.3$\pm$1.3* &  92.3$\pm$1.9*! &   88.7$\pm$3.6* &  80.9$\pm$10.8*! \\
DeepSDF   &   91.8$\pm$0.8* &  95.5$\pm$1.2*! &  94.3$\pm$1.3*! &  89.1$\pm$2.8*! &  93.3$\pm$1.5* &  92.1$\pm$2.0*! &  88.3$\pm$4.6*! &  80.0$\pm$11.2*! \\
\bottomrule
\end{tabular}
}
\end{table}

After examining the learned CHD type representation, we evaluated the ability of our method to represent existing CHD anatomies in the training dataset and to reconstruct previously unseen CHD anatomies in the testing dataset. Table \ref{table:seen} displays the Dice scores for the whole heart and the individual cardiac structures represented by different methods for the shapes seen in the training dataset. Adding correction significantly improved the accuracy of shape representation for our method, especially for the pulmonary artery, aorta, and atrium, which included the pulmonary veins and vena cavae. Indeed, the anatomies of blood vessels possess significant variations even within the same CHD types. Addressing these additional topological variations presented a challenge for the deformation network, necessitating the use of correction fields. Our method, with the correction network, achieved similar accuracy to CDeepSDF and DeepSDF. Without correction, our method achieved similar accuracy with NDF, which also represented shapes by deforming an implicit template. 

\begin{table}[h!]
\caption{Dice scores for whole heart (WH) and individual cardiac structures in reconstructed geometries compared to different baselines for unseen cases.(* denotes significance differences in mean values when compared with ``SDF4CHD''; denotes significance differences in mean values when compared with ``SDF4CHD+Corr''; we assume p-Values$< 0.0125 = 0.05/4$, based on a Bonferroni correction for four comparisons.)}
\label{table:unseen}
\resizebox{\textwidth}{!}{%
\begin{tabular}{lllllllll}
\toprule
{} &             WH &            LV &              RV &             LA &              RA &            Myo &              Ao &              PA \\
\midrule
SDF4CHD      &   85.9$\pm$2.9 &  91.1$\pm$2.5 &    87.7$\pm$8.8 &   82.1$\pm$6.7 &    88.9$\pm$3.2 &   89.0$\pm$4.9 &   74.4$\pm$13.6 &   70.2$\pm$17.1 \\
SDF4CHD+Corr &  88.5$\pm$2.2* &  91.1$\pm$2.6 &    87.9$\pm$8.1 &  85.5$\pm$5.8* &   92.0$\pm$2.5* &  90.4$\pm$3.6* &  80.4$\pm$10.3* &  73.6$\pm$18.2* \\
NDF       &  86.8$\pm$2.7! &  91.7$\pm$2.1 &    88.1$\pm$8.3 &  83.6$\pm$6.5! &   89.7$\pm$3.5! &  89.7$\pm$4.2! &  74.8$\pm$15.6! &   71.7$\pm$17.3 \\
CDeepSDF  &  88.5$\pm$2.2* &  92.0$\pm$1.9 &  91.1$\pm$3.4*! &  84.4$\pm$6.1* &  91.0$\pm$2.8*! &  90.4$\pm$3.7* &   82.7$\pm$7.7* &  71.4$\pm$19.1! \\
DeepSDF   &  88.5$\pm$2.3* &  92.1$\pm$1.9 &  91.0$\pm$3.6*! &  84.3$\pm$6.0* &  91.1$\pm$2.7*! &  90.3$\pm$3.9* &   83.0$\pm$7.5* &   72.0$\pm$18.3 \\
\bottomrule
\end{tabular}
}
\end{table}

To reconstruct unseen shapes, we utilized Maximum-a-Posterior (MAP) estimations of the shape code $Z_s$. Namely, for all methods, the unseen shapes were obtained by optimizing the shape code (and the correction code for our method) to minimize the objection function described by Equation \ref{eq:total} while freezing the parameters of the trained network. We initialized the shape code for each unseen shape as a $4 \times 4 \times 4$ tensor with random numbers sampled from Gaussian distribution, $\mathcal{N}(0, 0.01)$, and optimized the shape code for 200 iterations using the Adam optimizer\cite{adam} with a learning rate of 0.01. Table \ref{table:unseen} compares the Dice scores for cardiac anatomies reconstructed for unseen shapes in the testing dataset among different methods. Similarly to representing seen shapes, our method, with correction, achieved a similar shape reconstruction performance compared to CDeepSDF and DeepSDF. Without correction, our method achieved similar reconstruction performance with NDF.

\begin{table}[h!]
\caption{Dice scores for whole heart (WH) and individual cardiac structures in reconstructed geometries compared to different baselines for unseen cases when increasing the latent shape code dimension from $4\times4\times4$ to $8\times8\times8$(* denotes significance differences in mean values when compared with ``SDF4CHD''; denotes significance differences in mean values when compared with ``SDF4CHD+Corr''; we assume p-Values$< 0.0125 = 0.05/4$, based on a Bonferroni Correction for 4 comparisons.).}
\label{table:unseen8}
\resizebox{\textwidth}{!}{%
\begin{tabular}{lllllllll}
\toprule
{} &             WH &              LV &             RV &              LA &              RA &             Myo &             Ao &              PA \\
\midrule
SDF4CHD      &   92.9$\pm$1.3 &    95.5$\pm$0.8 &   94.1$\pm$3.2 &    90.4$\pm$3.9 &    95.1$\pm$1.2 &    93.9$\pm$2.2 &   88.1$\pm$7.0 &   81.6$\pm$15.0 \\
SDF4CHD+Corr &  94.2$\pm$1.0* &    95.5$\pm$0.8 &   94.3$\pm$2.3 &   92.7$\pm$3.2* &   96.5$\pm$0.8* &   94.3$\pm$1.9* &  91.2$\pm$5.0* &  85.5$\pm$12.3* \\
NDF       &  93.1$\pm$1.5! &    95.6$\pm$1.0 &   94.5$\pm$2.4 &  91.1$\pm$3.8*! &   95.4$\pm$1.4! &  94.1$\pm$2.0*! &  87.6$\pm$8.0! &  82.2$\pm$16.6! \\
CDeepSDF  &  93.1$\pm$1.2! &  95.2$\pm$0.8*! &   94.4$\pm$1.3 &   90.6$\pm$4.0! &   95.4$\pm$1.0! &  93.5$\pm$2.0*! &  91.5$\pm$2.1* &  82.1$\pm$13.6! \\
DeepSDF   &  93.0$\pm$1.1! &  95.1$\pm$0.9*! &  91.4$\pm$17.0 &   90.5$\pm$3.8! &  95.5$\pm$0.8*! &  93.4$\pm$2.1*! &  91.5$\pm$2.1* &  82.9$\pm$12.1! \\
\bottomrule
\end{tabular}
}
\end{table}

We observed that increasing the dimension of latent shape code when reconstructing unseen shapes effectively improved the reconstruction accuracy for all methods. Table \ref{table:unseen8} presents the Dice scores for various methods when reconstructing unseen shapes, following the increase in latent code dimensions from $4\times4\times4$ to $8\times8\times8$. This increase in spatial dimensionality expanded the shape representation capacity since the input shape code was tri-linearly sampled at spatial point locations to predict deformations (for our SDF4CHD method and NDF) or signed distance field values (for CDeepSDF and DeepSDF). In fact, when using an $8\times8\times8$ shape code size during inference, our method and others achieved higher average Dice scores on unseen shapes compared to using a $4\times4\times4$ shape code size on training shapes. 

\begin{figure}[h!]
    \centering
    \includegraphics[width=\linewidth]{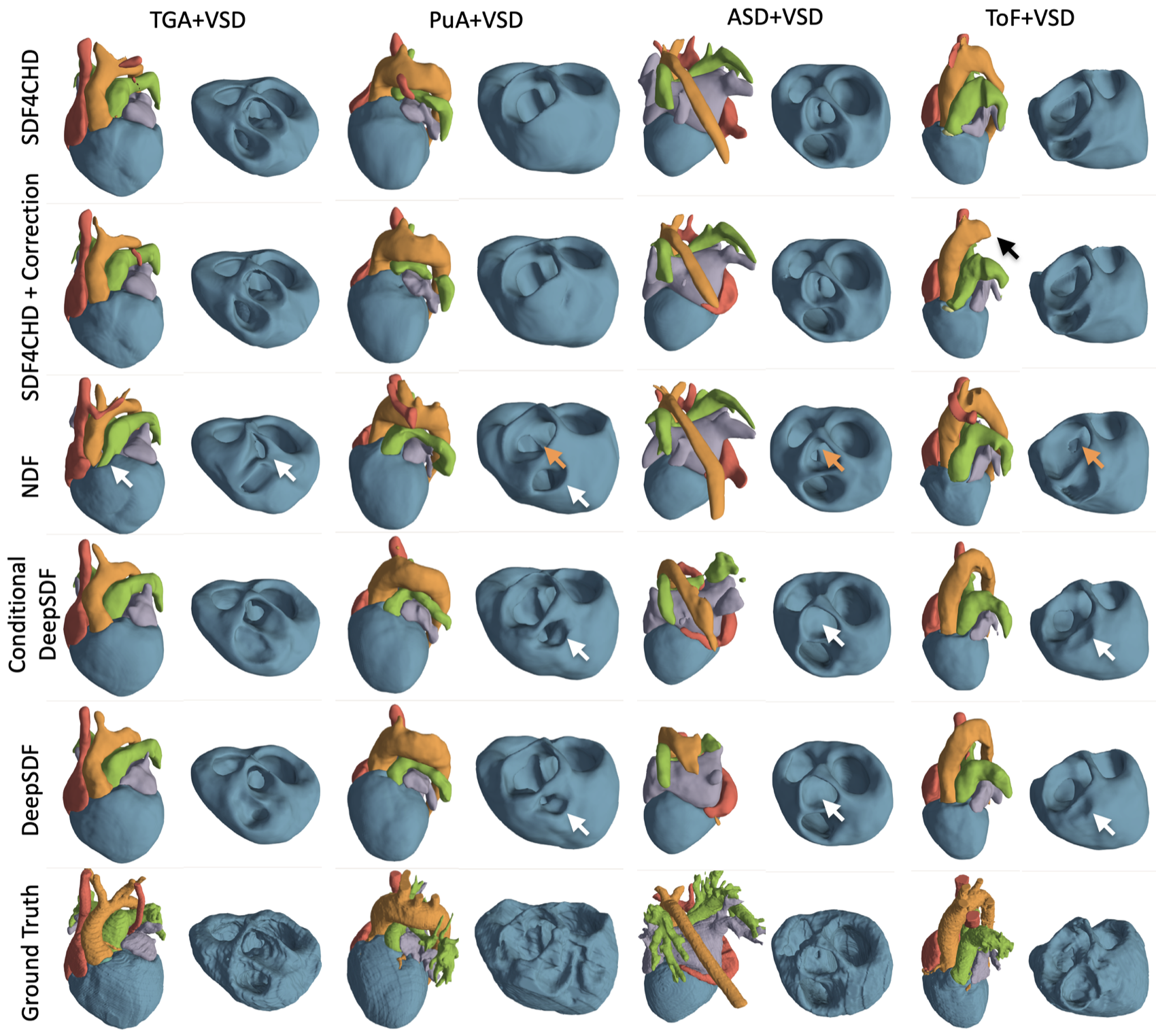}
    \caption{Example reconstruction results for unseen cases. The first and the second columns show the whole heart (WH) and myocardium (Myo) reconstruction, respectively. The white arrows indicate topological errors produced by NDF, Conditional DeepSDF, and DeepSDF. The orange arrows highlight that NDF sometimes failed to reconstruct accurate VSD sizes and geometries. The black arrow indicates an example where our correction module removed part of the aorta when the deformation module placed the aorta at inaccurate locations. }
    \label{fig:compare}
\end{figure}

In addition to demonstrating similar accuracy compared to previous approaches, our method offers a distinct advantage in producing cardiac anatomies that align with CHD diagnoses, as illustrated in Figure \ref{fig:compare}, which compares example reconstruction results for unseen cases. The limitation of the NDF approach lies in its inability to handle topological changes due to its reliance on a single implicit template. Given that a majority of its training cases feature VSDs, the implicit template learned by NDF consistently includes a VSD, rendering it unsuitable for reconstructing cardiac anatomies without VSDs. DeepSDF and CDeepSDF, can both model arbitrary topologies, but sometimes produce inconsistent topology. For example, these methods incorrectly predicted a pulmonary outflow tract in PuA where the pulmonary artery was supposed to disconnect from the right ventricle. They also did not correctly reconstruct smaller VSDs. In contrast, our method, by deforming a learned CHD type-specific implicit template that corresponded to the disease diagnosis, consistently reconstructed CHD anatomies that matched the diagnosis. Our use of a correction network was able to correct inaccuracies, mostly in the reconstruction of blood vessels. However, we observed that it sometimes removed half of the aorta if the deformation network failed to capture the correct location of the aorta. 

\begin{figure}[h!]
    \centering
    \includegraphics[width=\linewidth]{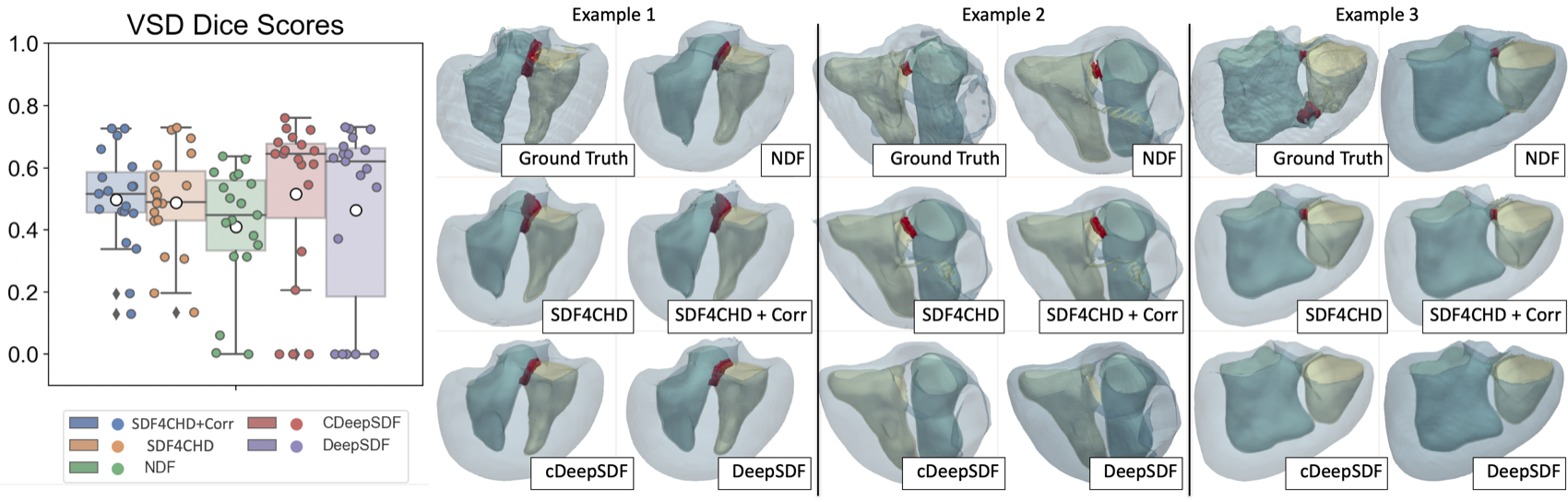}
    \caption{Left: Box plots showing Dice scores for VSDs (n=20) in the test dataset. The white circle shows the mean Dice score for each method, and the colored circles show the Dice scores of all data points. Right: example VSD reconstructions from different methods. The red regions represent the locations of VSDs where the RV blood pool and the LV blood pool meet. }
    \label{fig:vsd}
\end{figure}

We further compared the ability of our SDF4CHD method and the baseline methods to reconstruct local VSDs. VSDs were identified in the reconstructed cardiac anatomies as regions where the RV blood pool connected with the LV blood pool. To achieve this, we converted the predicted signed distance fields into segmentations and identified regions where the RV blood pool segmentation could be expanded to match the LV blood pool segmentation, and vice versa. Figure \ref{fig:vsd} (left) presents a comparison of the Dice scores for VSDs in a total of 20 reconstructed cardiac anatomies with a VSD diagnosis from the unseen test dataset. Our method, with or without corrections, effectively reconstructed anatomies with VSDs. In contrast, CDeepSDF and DeepSDF failed to reconstruct any VSDs in four and six cases, respectively. Since NDF deformed a learned implicit template with a VSD, its reconstructed cardiac anatomies all contained a VSD. The zero Dice score for NDF resulted from its inability to accurately deform the VSD in the template to match the correct location for the unseen anatomy. Figure \ref{fig:vsd} (right) presents three example cases, including one with a large VSD, another with a small VSD, and a case with two VSDs. In the case of the large VSD, all methods successfully produced reconstructions that included the VSD. However, when dealing with the small VSD, both CDeepSDF and DeepSDF failed to detect this defect. We note that accurately reconstructing small VSDs when training on global cardiac anatomies poses a considerable challenge. Our method, leveraging diagnosis priors and CHD type-specific templates, demonstrated more robust performance in capturing these local defects. Nevertheless, our learned VSD template anatomies only contained a subaortic VSD, which led to our method's failure in reconstructing the apical VSD in the case with both VSDs present. Notably, the correction network didn't capture the apical VSD, and the baseline methods also failed in this regard. This specific defect was only present in this single test case, while all training shapes featured subaortic VSDs, causing all trained methods to struggle with this out-of-distribution defect.

\subsection{Dataset Generation}
\begin{figure}[h!]
    \centering
    \includegraphics[width=\linewidth]{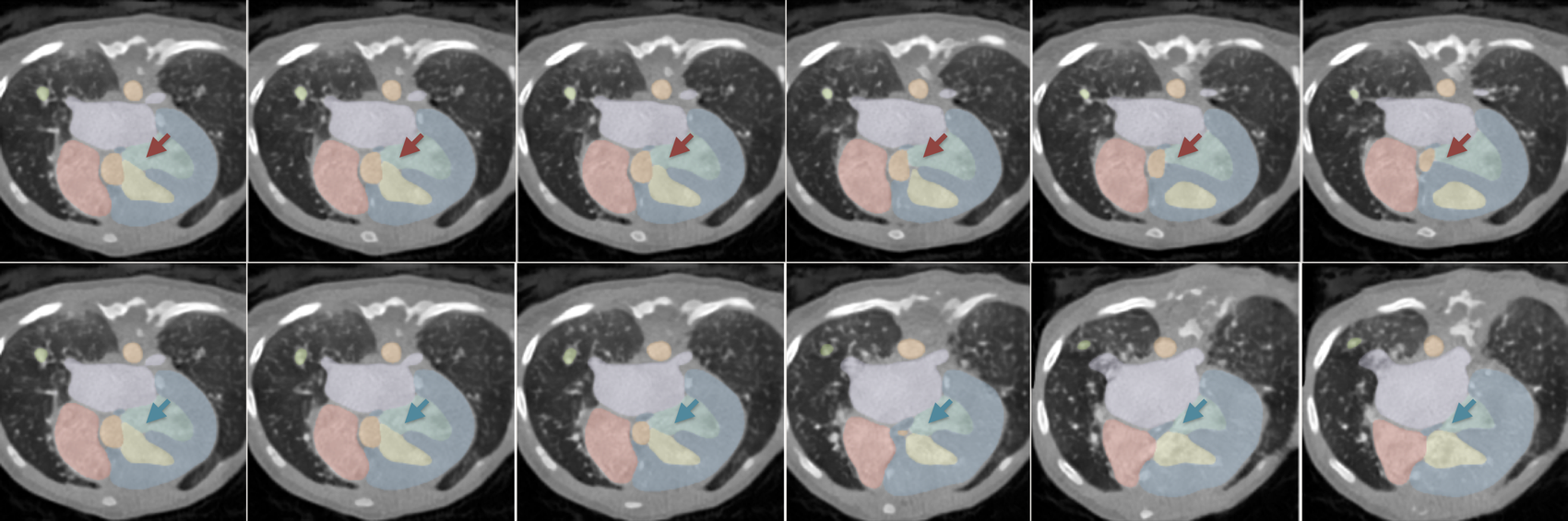}
    \caption{Example synthetic CT image and segmentation pairs. The top row displays the synthetic data when simulating a CHD-type change from the disease state of PuA and VSD to a healthy state. The bottom row displays the synthetic data for a CHD-shape change, that is, deforming a patient-specific geometry to the shape of a healthy patient while preserving the CHD type (PuA and VSD). The red arrows highlight the gradual shrinking of VSD from disease to a healthy state, whereas the blue arrows highlight the preservation of VSD during shape changes. }
    \label{fig:syn_img}
\end{figure}

Our method can be applied to generate a spectrum of CHD datasets for segmentation and computational simulations. For example, we can employ a conditional Generative Adversarial Network (cGAN) \cite{cGAN2014} to synthesize paired CT images from the generated anatomies. The synthesized images match with the segmentations created from our generative anatomies, and can thus expand our dataset on underrepresented and uncommon CHD types. Such an expanded dataset can be used for training a neural network to improve CHD segmentation performance.  Figure \ref{fig:syn_img} illustrates the synthesized image-segmentation pairs representing intermediate states between ToF and PuA for patient-specific anatomies. It also showcases deformations between two patient-specific shapes, with the CHD type held constant. Specifically, we followed the implementation of Amirrajab \etal \cite{Amirrajab_2023}, which leverages specially designed normalization, known as Spatially Adaptive DEnormalisation (SPADE) \cite{SPADE}. Using this approach, we successfully preserved tissue boundaries described in the input conditional segmentation and synthesized stacks of 2D CT images with an appearance similar to a reference CT image.

\begin{figure}[h!]
    \centering
    \includegraphics[width=\linewidth]{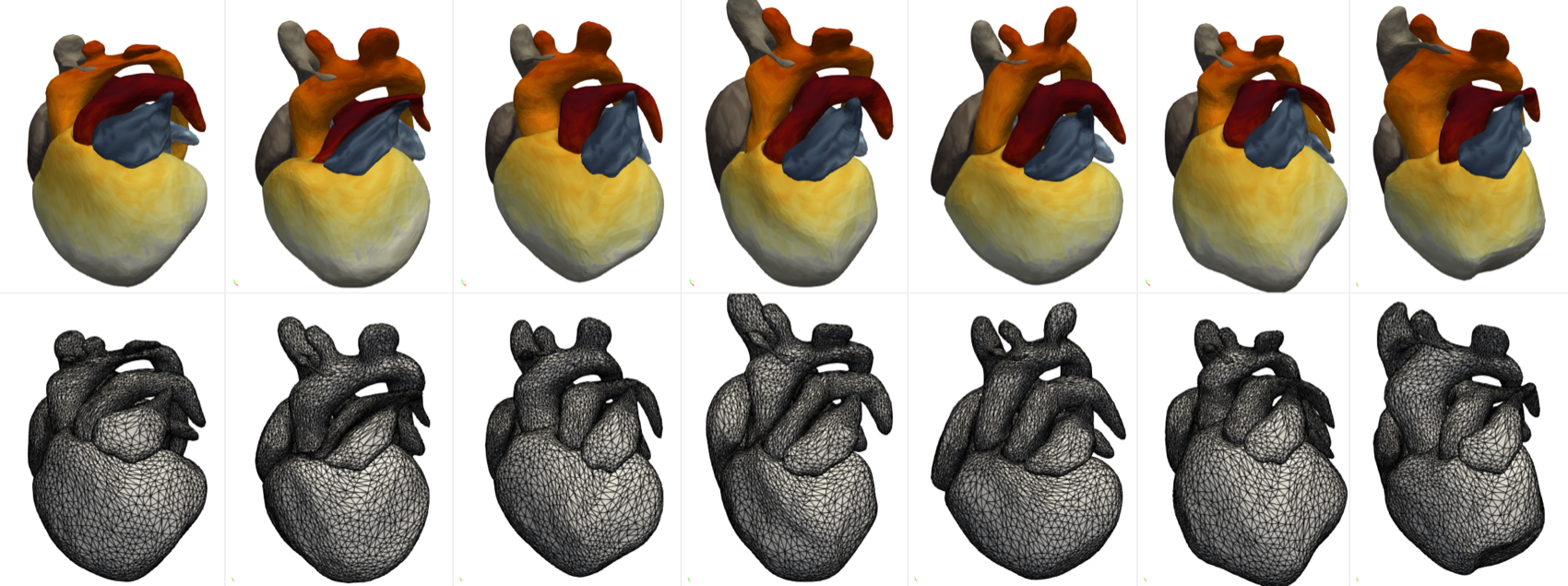}
    \caption{Cardiac meshes generated by deforming a DORV-type-specific mesh template. Textures were mapped based on mesh vertex IDs showing semantic correspondence among generated meshes.}
    \label{fig:syn_mesh}
\end{figure}

Furthermore, our method can generate deformed meshes of specific CHDs for simulations or 3D printing (Figure \ref{fig:syn_mesh}. A template mesh can be created from the type-specific SDF and then deformed in a diffeomorphic manner, thanks to our use of NODE. The generated meshes within a CHD type have semantic correspondence, as indicated by the texture map that is created based on global vertex IDs, and which shares the same connectivity as the type-specific template mesh. 

\subsection{Analysis of Design Choices}
%
    
We conducted ablation analyses to understand the effects of several design choices on our model's performance. Namely, we considered the effect of alternatively training type and shape networks, using Lipschitz regularization on the type network, using latent codes of different dimensions, and using NODE rather than direct displacement prediction to deform the learned CHD type anatomies. 

\begin{figure}[h!]
    \centering
    \includegraphics[width=\linewidth]{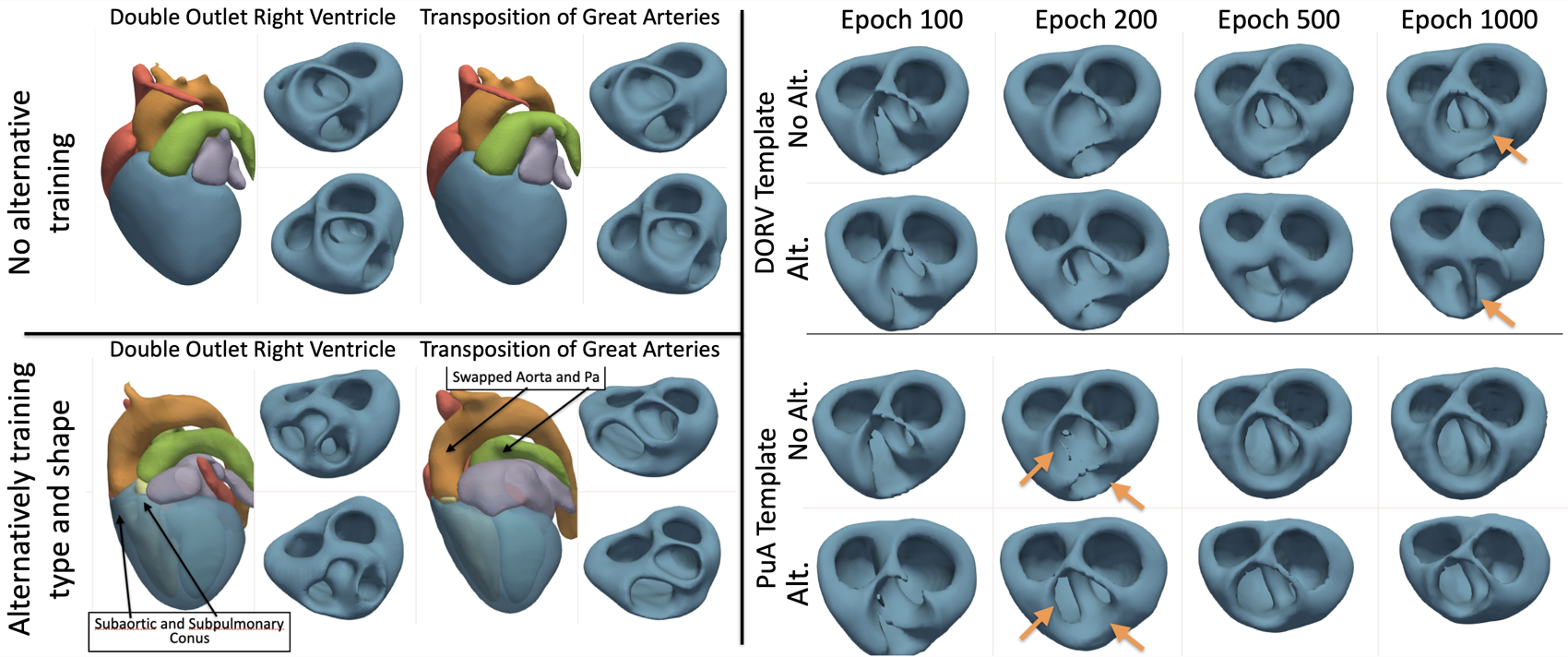}
    \caption{The resulting learned CHD type-specific templates obtained from joint training and from alternative training the shape and type networks, respectively. Left: The final templates for DORV and TGA. Right: The templates for DORV and PuA after various epochs during training. Orange arrows highlight CHD abnormalities that were better captured by using alternative training. }
    \label{fig:alter_train}
\end{figure}

\begin{figure}[h!]
    \centering
    \includegraphics[width=0.6\linewidth]{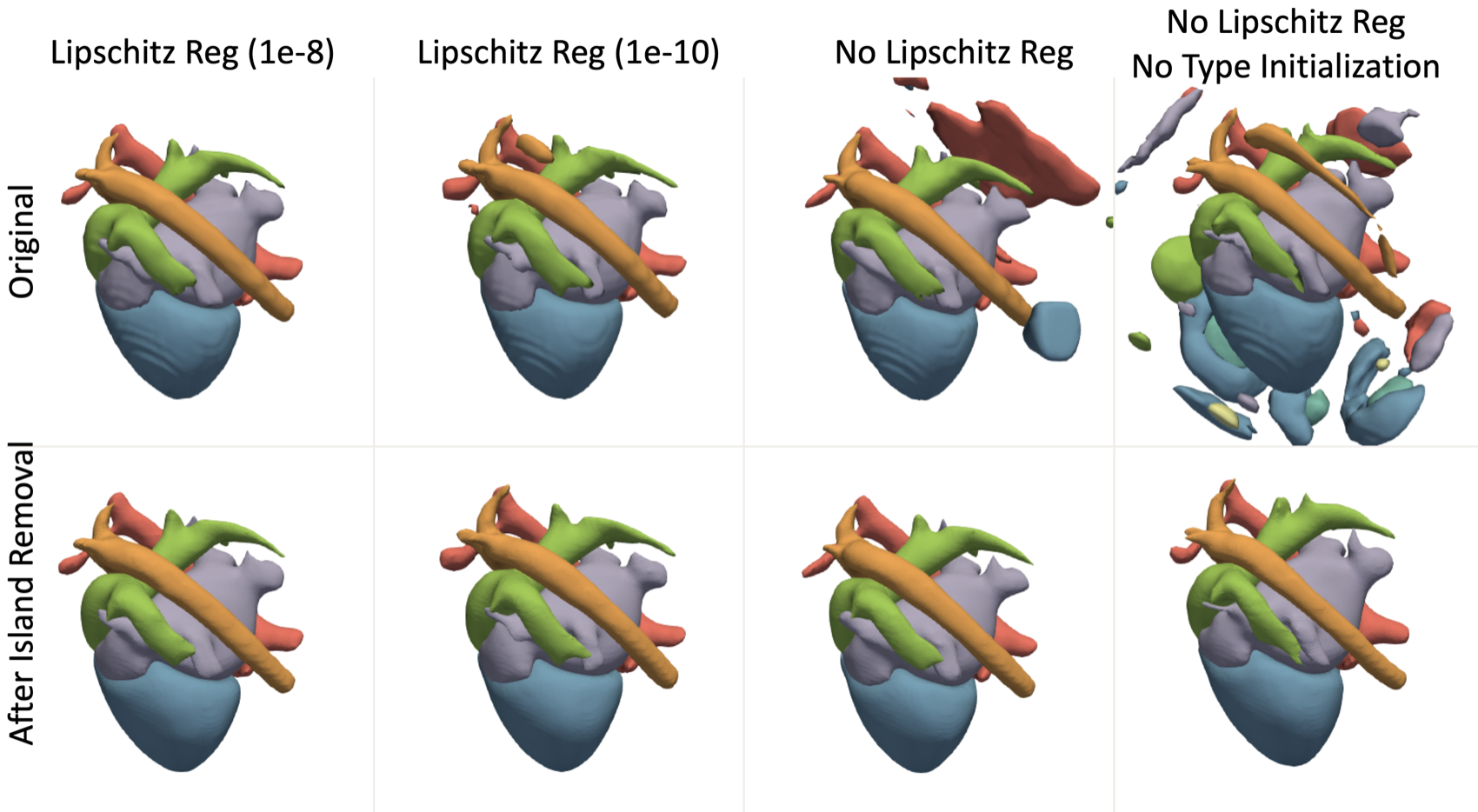}
    \caption{A comparison among the resulting CHD type-specific templates learned from training the type network with Lipschitz regularization ($\lambda_{lip}=1e^{-8}$, $\lambda_{lip}=1e^{-10}$), without Lipschitz regularization, without Lipschitz regularization or type initialization. The top row shows the original learned template, and the bottom row shows the corresponding template after post-processing (i.e. extracting the largest connected components.) }
    \label{fig:lipreg}
\end{figure}

Alternatively training the type and shape network helped our method learn the CHD type representations faster and more robustly. While joint training was able to correctly learn the CHD type-specific anatomies for ToF, PuA, VSD, and normal cases, it failed to correctly represent the position changes between the aorta and the pulmonary arteries for DORV and TGA, as demonstrated in Figure \ref{fig:alter_train} (left). Furthermore, alternative training was able to learn the large VSD and the discontinued pulmonary artery in PuA before Epoch 200, whereas joint training took longer to capture these anatomical abnormalities. Furthermore, we observed that applying Lipschitz regularization to the type network and initializing the type network by pre-training it to reconstruct healthy cardiac anatomy, regardless of the input diagnosis vector, can reduce the presence of noisy isolated islands in the learned CHD type-specific anatomies, as demonstrated in Figure \ref{fig:lipreg}. However, such artifacts can be easily corrected by a post-processing step where we only kept the largest connected component of a cardiac structure. We did not observe any notable effects of Lipschitz regularization and initialization when interpolating among different CHD type-specific templates. 

We observed that the spatial resolution (i.e. dimension) of latent shape codes significantly affects the reconstruction accuracy. Specifically, higher-dimensional or spatially-resolved shape codes exhibit greater precision in reconstructing previously unseen geometries. This is because, when we perform linear interpolation on cubical shape codes, a higher spatial resolution results in more distinct values among the shape latent vectors at closely located points. Consequently, this encourages the deformation network $\mathcal{D}$ to capture finer details. We noticed that the spatial resolution of the latent shape code during training had minimal impact on reconstruction accuracy. Specifically, training with a low-resolution shape code of $2^3$ produced similar accuracy as training with a high-resolution shape code of $8^3$. However, a high-resolution shape code would require a larger dataset to effectively learn a latent space exclusively comprising physiological shapes. Notably, when generating synthetic cardiac anatomies, an $8^3$ high-resolution shape code resulted in non-physiological distortions. Therefore, we chose a latent dimension of four for our shape codes to achieve a balance between reconstruction and generation performance.

\begin{figure}[h!]
    \centering
    \includegraphics[width=\linewidth]{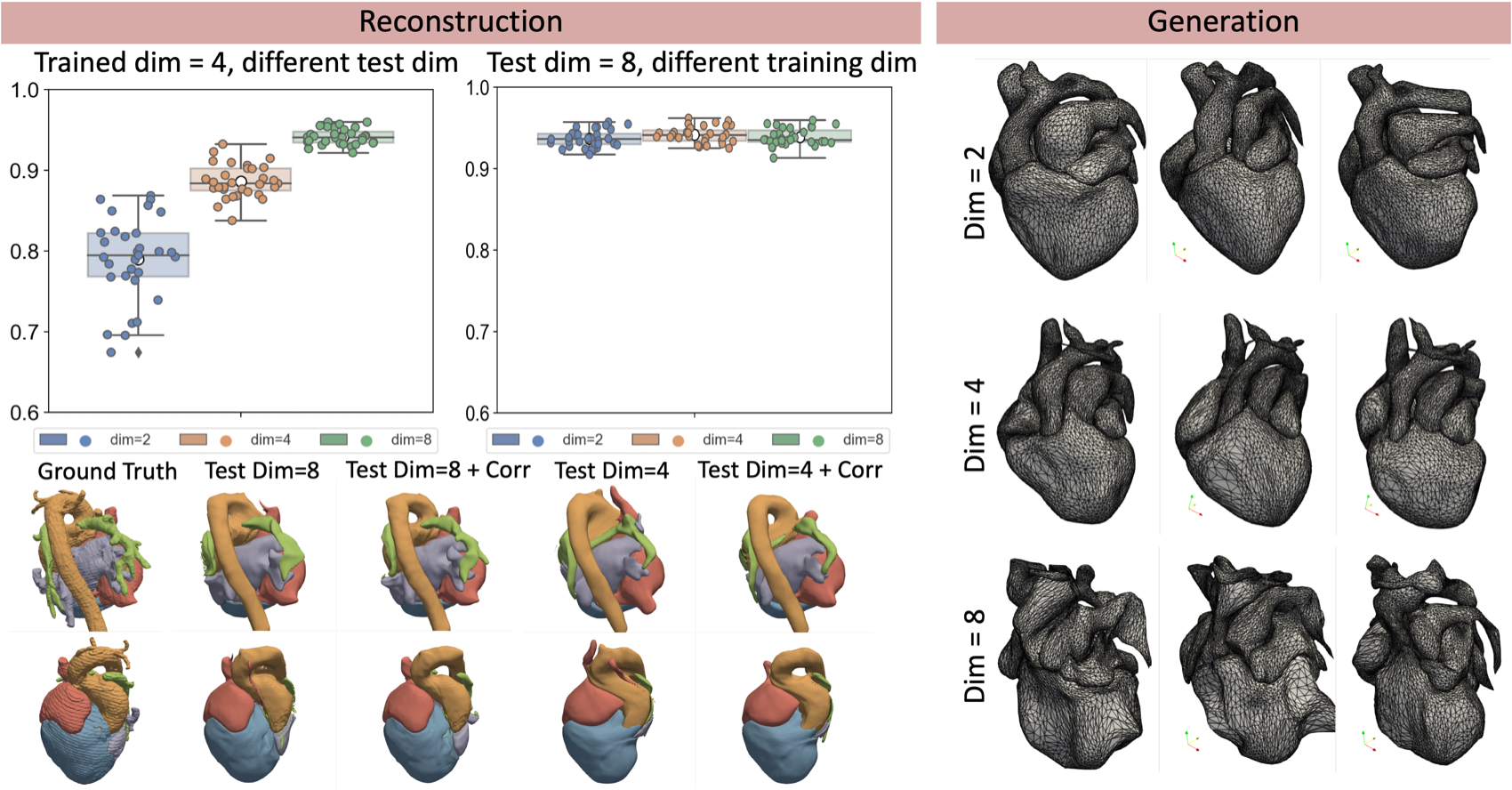}
    \caption{Left: Effect of latent shape code dimension on cardiac shape reconstruction. The top panel shows box plots of whole heart Dice scores evaluated on the test set when testing with latent shape code of various dimensions for a model trained on a latent shape code dimension of $4^3$, and when testing with the same latent shape code dimension of $8^3$ after being trained with latent shape codes of various dimensions. The bottom left panel shows a qualitative comparison between reconstructing shapes using shape code dimensions of $8^3$ and $4^3$ (with or without the correction module), respectively. Right: Effect of latent shape code dimension on cardiac shape generation where random shape codes of different dimensions ($2^3$, $4^3$, $8^3$) were sampled from Gaussian prior distributions of $\mathcal{N}(0, 0.01^2I)$ to generate synthetic cardiac anatomies.}
    \label{fig:latent_dim}
\end{figure}

\begin{figure}[h!]
    \centering
    \includegraphics[width=0.6\linewidth]{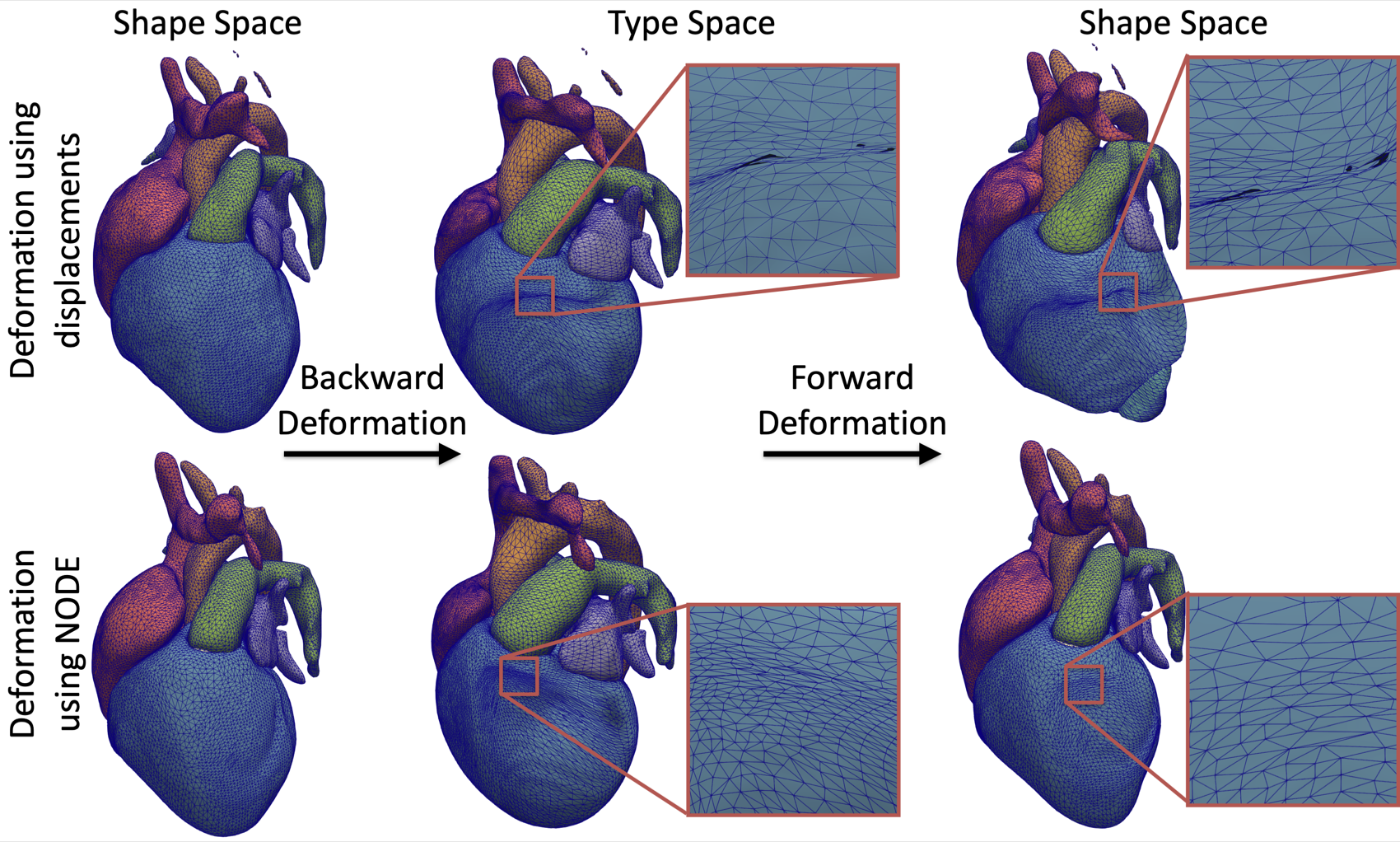}
    \caption{Comparison between predicting an invertible flow field (bottom) and predicting displacements directly (top) to deform shapes. From left to right shows the patient-specific meshes extracted from the signed-distance field at the shape space, the CHD type-specific meshes obtained by deforming the patient-specific mesh to the type space, and the reconstructed patient-specific meshes obtained by deforming the CHD type-specific mesh back to the shape space. The zoomed-in panels highlight that by using a NODE, we prevented the mesh intersections resulting from directly predicting the displacements.}
    \label{fig:sif}
\end{figure}

Figure \ref{fig:sif} compares two methods of deforming cardiac anatomies. The first one is our currently adopted approach of using NODE to learn an invertible flow field, whereas the second one is directly predicting displacements on mesh vertices, as adopted in our patient-specific mesh-reconstruction approach\cite{Kong2022LearningWH}. As Figure \ref{fig:sif} shows, since the networks were trained to deform from shape to type space, learning an invertible flow enabled our method to readily invert the deformation and deform the mesh from type to shape space during test time to generate patient-specific meshes, whereas directly predicting displacements generated inaccurate results. Furthermore, deformation under an invertible flow prevents intersections of point trajectories, thereby preventing inverted mesh elements. We used TetGen \cite{Si2013TetGenAQ} to detect surface intersections on synthetic whole-heart meshes, each of which contained about 20K mesh vertices. Table \ref{table:sif} shows that predicting the displacements produced surface self-intersections, whereas using NODE produced 0 intersections. 

\begin{table}[h!]
\caption{Percentage of intersected mesh elements for synthetic meshes. The meshes were generated by deforming CHD type-specific template meshes, either using predicted invertible flow fields (NODE) or using directly predicted displacements (Displacement) on mesh vertices.}
\label{table:sif}
\resizebox{\textwidth}{!}{%
\begin{tabular}{llllllll}
\toprule
{} &             LV &             RV &            LA &             RA &            Myo &             Ao &             PA \\
\midrule
NODE         &    0.0$\pm$0.0 &    0.0$\pm$0.0 &   0.0$\pm$0.0 &    0.0$\pm$0.0 &    0.0$\pm$0.0 &    0.0$\pm$0.0 &    0.0$\pm$0.0 \\
Displacement &  0.22$\pm$0.37 &  0.28$\pm$0.37 &  0.2$\pm$0.24 &  0.18$\pm$0.23 &  0.16$\pm$0.24 &  0.47$\pm$0.59 &  0.26$\pm$0.23 \\
\bottomrule
\end{tabular}
}
\end{table}

\section{Discussions}

CHDs are uncommon conditions that can significantly impact a patient's cardiac function, often necessitating specialized expertise for diagnosis, treatment planning, and surgical interventions. Computer-assisted methods, such as 3D printing, visualization, image analysis, and, computational simulations, hold promise in improving CHD patient care. Nonetheless, due to the rarity of CHDs, there exists a significant challenge in assembling adequate patient cohorts and associated ground truth data to fully harness the potential and validate these techniques. Generative DL methods can model the distribution of cardiac anatomies from a training dataset. They can thus create an unlimited number of synthetic anatomies by drawing from the learned distribution. However, prior generative DL methods are only suitable for normal cardiac anatomies and cannot model the diverse anatomical abnormalities seen in CHDs.  We thus develop a novel generative DL approach that represents CHD cardiac anatomies in a CHD-type and CHD-shape disentangled manner. Leveraging readily available CHD diagnosis information, our approach uses an SDF-based type-representation module to model the corresponding type-specific cardiac abnormalities and uses deformation and correction modules to represent patient-specific variations of cardiac shapes. Our method uniquely enables type-controlled representation, generation, and reconstruction of cardiac anatomies for a range of complex CHDs.

Our method introduces topological variations in the representation of cardiac anatomies, an important and previously unaddressed consideration for modeling CHD. Previous studies have largely focused on cardiac diseases such as hypertrophic cardiomyopathy \cite{Biffi2019ExplainableAS}, myocardial infarction \cite{Beetz2022InterpretableCA}, and dilated cardiomyopathy \cite{Gmez2021ADC}, where the topology and arrangement of cardiac structures typically remain conventional, with abnormalities primarily confined to changes in shape. Furthermore, our CHD-type and CHD-shape disentangled representation uniquely supports CHD type-controlled generation of cardiac anatomies for downstream applications, such as image segmentation and computational simulations tailored to specific CHD cases. Our method achieves this using a single trained network, which automatically learns generic anatomical templates corresponding to different CHDs. In contrast, previous methodologies, such as NDF \cite{Sun2022TopologyPreservingSR} and HeartDeformNet \cite{Kong2022LearningWH}, assume a single topology, which may require separate dataset training or manual design of distinct mesh templates for different CHDs, respectively. By deforming the learned generic templates, our method also exhibited advantages in reconstructing patient cardiac anatomies with improved consistency with the CHD diagnosis. 

We demonstrate the effectiveness of implicit SDF-based anatomy representation in modeling CHD hearts. Specifically, it empowers us to model the diverse topologies and arrangements of cardiac structures in CHDs while representing intermediate CHD states beyond the scope of binary diagnoses. Although CHD diagnoses are typically viewed as categorical, it's important to acknowledge that CHD anatomies encompass a wide and continuous spectrum of topological variations across different defect types originating from abnormal fetal heart development. A Lipchitz-regularized SDF can effectively capture gradual and smooth changes in anatomical abnormalities. We have shown that when linearly interpolating the input CHD-type vector, our method can accurately represent the anatomical changes that align with intermediate diagnoses. The ability to handle potential CHD topological variations, even with limited training data, is valuable in addressing the challenge of acquiring diverse CHD types for training data due to their extreme rarity.

We note that our method captures both the topological abnormality and the shape abnormality in its type representation module. For example, our type representation for ToF captured the thickening of the right ventricular myocardium and the narrowing of the pulmonary outflow tract. Here, the shape variations captured in the type representation module are responsible for characteristic shape features of the particular disease types, whereas the shape variations captured in the shape latent codes and the deformation network are (assumed to be) independent of diseases. Our ability to capture both the CHD-shape and the CHD-type abnormalities for specific cardiac diseases originates from, perhaps counterintuitively, the CHD-type and CHD-shape disentangled design. Namely, our deformation network morphs the generic CHD-type templates to match the cardiac anatomies for different diseases in the training dataset. Since the prior distribution of the shape codes is assumed to follow the Gaussian distribution independent of CHD types, the deformation network is biased towards only modeling the patient-specific shape variations that are not due to CHDs. Therefore, our CHD-type representation module is forced to capture the typical shape and type abnormalities for each disease type in the training dataset. 

An immediate advantage of the CHD-type and shape disentangled design is that the represented generic CHD-type anatomies for different CHD types are aligned in space, and the differences in their anatomies are primarily attributed to the corresponding abnormalities for the different CHDs. Therefore, by applying the same patient-specific shape code and changing the type code, we can readily visualize the anatomical changes corresponding to various degrees of defects in the same patient-specific cardiac geometries. This could also benefit clinical education, since the current education is mostly based on visualizations of 3D patient-specific geometries, confounding the CHD type-specific variations with the patient-specific shape variations. Furthermore, our type encoding module can potentially represent anatomy changes before and after surgical repairs, for example, the insertion of intracardiac patches to close VSD or septate the heart in DORV \cite{BACKER2003121},  and the arterial switch operation to swap the pulmonary artery and aorta positions in TGA \cite{Quinn2008TheML}. Therefore, our method can potentially be applied to visualize and examine the cardiac anatomies following surgical operations for specific patients, thereby facilitating surgical planning. 

We generated the whole heart cardiac meshes by extracting the zero-level-set surface mesh of the learned generic type-specific SDF and deforming the mesh differently using shape codes randomly sampled from a Gaussian prior distribution. However, since the deformation network acts on any points in the space, one can also readily deform a different mesh adapted from the zero-level-set surface mesh for various computational applications. For example, one could edit the template mesh to create simulation-ready mesh templates so that following deformation, the resulting meshes can readily be used for computational simulations of electrophysiology, active contraction, and fluid dynamics within the heart. Compared with our prior approach that deforms a template mesh by directly predicting displacements, we demonstrated that using a NODE approach to deform a mesh in a diffeomorphic manner can effectively reduce mesh artifacts such as surface intersections. However, the predicted diffeomorphic mapping does not guarantee that the deformed mesh elements will have strictly positive Jacobian values, since the learned diffeomorphism can contain sinks, making the points nearby converge to the same location. This can be addressed by including a regularization loss developed by \cite{Narayanan2023}, which penalizes vanishing volumes during training and test-time optimization. 

We assumed that the shape codes follow the same Gaussian prior distribution for all CHD types. This is a simplifying assumption considering the limited training dataset that we currently have. Indeed, although the CHD type-representation module can capture typical CHD-related shape abnormalities, the patient-specific shape distributions of cardiac anatomies may still differ for different CHD types. For example, the pulmonary arteries of patients with PuA pose a much wider range of variations compared with normal patients in terms of the numbers, locations, and orientations of major branches. However, it is challenging to fully capture such variations in shape distributions using only a few training samples for each CHD type. Nevertheless, if a larger dataset for CHD patients such as \cite{Xu2023ACA} can be made available, our method can easily be adapted to learn CHD type-conditioned shape distributions. Namely, Biffi \etal \cite{Biffi2019ExplainableAS} learned the shape latent distribution differences between HCM and healthy cases by training another network to discriminate HCM shape codes from healthy shape codes. We can adapt this approach to perform CHD-type classification tasks in the latent shape code space. This will result in a well-separated latent shape code distribution for different CHD types. Our method can thus be extended for analyzing shape variation differences among various CHD types in the future.


\section{Conclusions}

We have introduced a novel deep-learning approach that learns a CHD-type and CHD-shape disentangled representation of cardiac geometry for major CHD types. Our approach implicitly represents type-specific anatomies of the heart using neural SDFs and learns an invertible deformation for representing patient-specific shapes. In contrast to prior generative modeling approaches designed for normal cardiac topology, our approach accurately captures the unique cardiac anatomical abnormalities corresponding to various CHDs and provides meaningful intermediate CHD states to represent a wide CHD spectrum. When provided with a CHD-type diagnosis, our approach can create synthetic cardiac anatomies with shape variations, all while retaining the specific abnormalities associated with that CHD type. We demonstrated the ability to augment image-segmentation pairs for rarer CHD types and generate cardiac meshes with consistent vertex connectivity. In the future, we plan to explore applying this approach to facilitate virtual surgery planning, image segmentation, classification, and computational simulations of cardiac function for CHD patients.

\section*{Acknowledgments}
This work was supported by NSF 1663671, NIH R01EB029362, NIH R01LM013120, and NIH R38HL143615. We thank Allyson Weiss for her assistance with creating ground truth segmentation.

\newpage

\newcommand{\beginsupplement}{
  \setcounter{table}{0}  
  \renewcommand{\thetable}{S\arabic{table}} 
  \setcounter{figure}{0} 
  \renewcommand{\thefigure}{S\arabic{figure}}
  \renewcommand{\thesubsection}{\Alph{subsection}}
  \counterwithin{figure}{subsection}
}
\beginsupplement

\section*{Supplemental Materials}

\subsection{Processing of ``imageCHD'' Dataset}

The ground truth segmentation provided by the ``imageCHD'' dataset \cite{Xu2020ImageCHDA3} had a few limitations. First, the images and segmentations provided did not contain slice spacing information, which caused some hearts in the original dataset to be un-physiologically flat along a certain axis. We assigned spacing values to the segmentation volumes so that a heart spans the same length when measured along axial, sagittal, and coronal axes for each volume. Moreover, the segmentations included numerous small vessel branches that were outside the scope of modeling cardiac defects. Therefore, we applied Gaussian smoothing with a standard deviation of 2 mm to eliminate these smaller vessel branches. Additionally, the myocardium segmentations in the dataset exhibited noise, with the presence of trabeculae structures and undesired holes and gaps near the atrium and great vessel openings, which were unrelated to CHDs. We manually corrected these artifacts, removing noise to ensure that the resulting myocardium segmentations aligned with the provided diagnosis and image data. We used the ``Segmentation Editor'' tool in 3D Slicer to process the segmentation. 

\begin{figure}[h!]
    \centering
    \includegraphics[width=\linewidth]{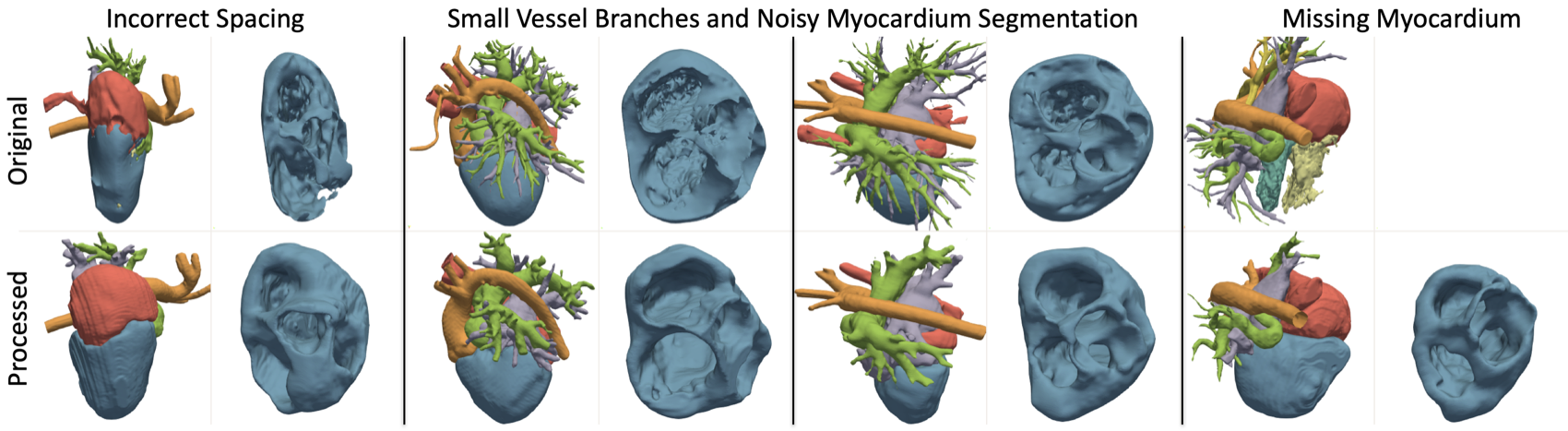}
    \caption{Comparison of ground truth segmentations provided by the ``ImageCHD" dataset and after manual processing.}
    \label{fig:ap:ground_truth}
\end{figure}

\subsection{Implementation and Optimization Details}

All neural networks in SDF4CHD were implemented using six fully connected layers with leaky ReLU activation functions ($\alpha = 0.02$). Each layer contains 512 neurons. We used six encoding functions\cite{Tancik2020FourierFL} for $\mathcal{T}_{dec}$ and four encoding functions for $\mathcal{D}$ and $\mathcal{C}$ to augment the input coordinate values.  

For DeepSDF, we utilized the default implementation and training configuration \cite{Park2019DeepSDFLC}, employing eight fully connected layers, each consisting of 512 neurons. We also used a residual connection from the input to the fourth layer. In the case of conditional DeepSDF, we adopted the same $\mathcal{T}_{enc}$ implementation as in SDF4CHD for encoding the diagnosis information, which was subsequently concatenated with the shape code. For NDF, we employed the same $\mathcal{T}_{dec}$ implementation as in SDF4CHD to predict the template SDF, which was then deformed using a deformation network implemented in the same manner as $\mathcal{D}$. Using the same network implementation ensured a fair comparison between the baselines and our method to verify the advantages of our CHD-type and shape disentangled design. Furthermore, we applied the same number of encoding functions, latent code dimensions, and training procedures for all methods to maintain a fair comparison.

\begin{figure}[h!]
    \centering
    \includegraphics[width=0.5\linewidth]{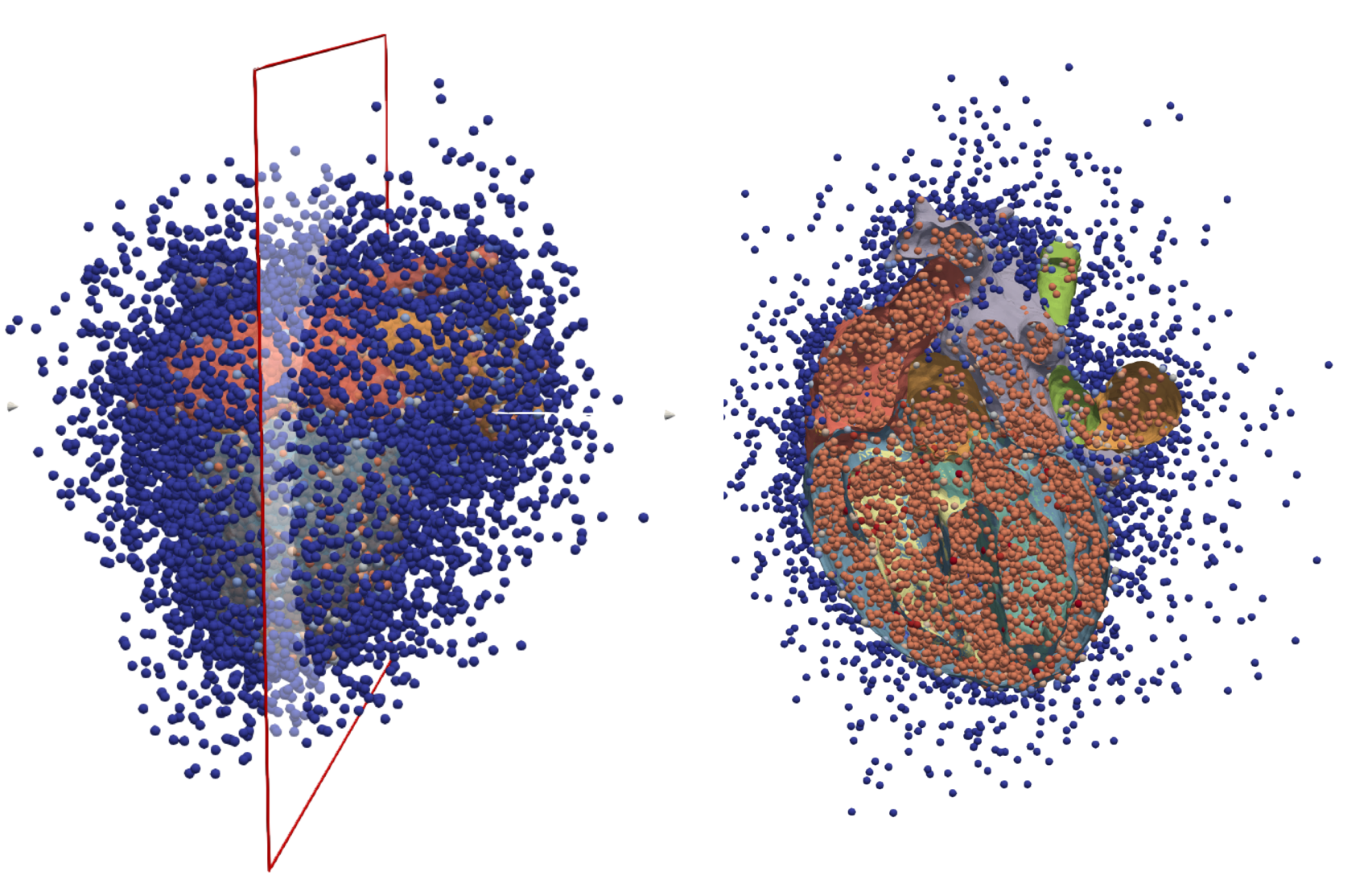}
    \caption{Example point cloud sampled during training. On the left, the point cloud is overlaid with ground truth whole heart surfaces. On the right, a cross-sectional view shows the point cloud and surfaces sliced at a plane. The blue points are outside the heart, while the orange points are inside the heart.}
    \label{fig:ap:sample}
\end{figure}

The network parameters, shape codes, and correction codes were computed by minimizing the total loss function using the Adam stochastic gradient descent algorithm \cite{adam}. The initial learning rate was set to be 0.0001, while $\beta_1$ and $\beta_2$ for the Adam algorithm were set to 0.5 and 0.999, respectively. We sampled 16384 points during each gradient descent update. The points were sampled from a probability distribution based on their distance from the boundaries of the cardiac structure, with the points located on the cardiac boundaries assigned the highest probability. Figure \ref{fig:ap:sample} shows an example point cloud sampled during a training step. We adopted a learning rate schedule where the learning rate was reduced by 30\% if the validation point losses had not improved for 20 epochs. The minimum learning rate was $1\mathrm{e}{-6}$. The network was implemented in PyTorch and the training was conducted on a Nvidia A40 graphics processing unit (GPU).

\bibliographystyle{unsrt}  
\bibliography{references}  
\end{document}